\documentclass[12pt,journal,onecolumn]{IEEEtran}
\usepackage{amsmath,amssymb, dsfont,bbm}
\usepackage{epsfig,latexsym,amssymb,amsmath,graphics,color}
\usepackage{graphicx,subfigure}
\usepackage{float}
\usepackage[linesnumbered,ruled]{algorithm2e}
\usepackage{epsfig,latexsym,amssymb,amsmath,graphics}
\usepackage{graphicx}
\usepackage[linesnumbered,ruled]{algorithm2e}
\usepackage{cite,color}
\usepackage{url,epstopdf}
\newtheorem{Proposition}{Proposition}
\newtheorem{theorem}{Theorem}
\newtheorem{lemma}{Lemma}
\newtheorem{remark}{Remark}
\newtheorem{proposition}{Proposition}
\newtheorem{corollary}{Corollary}

\newtheorem{prop}{Proposition}

\newtheorem{cor}{Corollary}

\newtheorem{lm}{Lemma}

\newtheorem{thm}{Theorem}

\newcommand{\be}{\begin{eqnarray}}
\newcommand{\ee}{\end{eqnarray}}
\newcommand{\benn}{\begin{eqnarray*}}
\newcommand{\eenn}{\end{eqnarray*}}
\def\IR{\rm I \kern-0.20em R}
\newcommand{\utwi}[1]{\mbox{\boldmath $ #1$}}

\newcommand{\bthm}{\begin{thm}}
\newcommand{\ethm}{\end{thm}}

\newcommand{\bcor}{\begin{cor}}
\newcommand{\ecor}{\end{cor}}
\newcommand{\bprop}{\begin{prop}}
\newcommand{\eprop}{\end{prop}}
\newcommand{\blm}{\begin{lm}}
\newcommand{\elm}{\end{lm}}
\newcommand{\beq}{\begin{equation}}
\newcommand{\eeq}{\end{equation}}
\newcommand{\ber}{\begin{eqnarray}}
\newcommand{\eer}{\end{eqnarray}}

\newcommand{\bproof}{\begin{proof}}
\newcommand{\eproof}{\end{proof}}

\newcommand{\bit}{\begin{itemize}}
\newcommand{\eit}{\end{itemize}}
\newcommand{\ben}{\begin{enumerate}}
\newcommand{\een}{\end{enumerate}}
\newcommand{\bdesc}{\begin{description}}
\newcommand{\edesc}{\end{description}}
\newcommand{\beqarrn}{\begin{eqnarray*}}
\newcommand{\eeqarrn}{\end{eqnarray*}}
\newcommand{\bproofof}{\begin{proofof}}
\newcommand{\eproofof}{\end{proofof}}
\newenvironment{rem}{\begin{trivlist}\item[]{\bf
Remark:}\hspace{4mm}}{\end{trivlist}}
\newcommand{\brem}{\begin{rem}}
\newcommand{\erem}{\end{rem}}
\newenvironment{rems}{\begin{trivlist}\item[]{\bf
Remarks}\begin{itemize}}{\end{itemize}\end{trivlist}}
\newcommand{\brems}{\begin{rems}}
\newcommand{\erems}{\end{rems}}
\newtheorem{fact}{Fact}
\newcommand{\bfact}{\begin{fact}}
\newcommand{\efact}{\end{fact}}
\newtheorem{examp}{Example}
\newcommand{\bexamp}{\begin{examp}\rm}
\newcommand{\eexamp}{\end{examp}}
\newtheorem{defn}{Definition}
\newcommand{\bdefn}{\begin{defn}\rm}
\newcommand{\edefn}{\end{defn}}

\newtheorem{alg}{Algorithm}
\newcommand{\balg}{\begin{alg}}
\newcommand{\ealg}{\end{alg}}

\newtheorem{prob}{Problem}
\newcommand{\bprob}{\begin{prob}}
\newcommand{\eprob}{\end{prob}}

\newcommand{\bvtm}{\begin{verbatim}}
\newcommand{\bfig}{\begin{figure}}
\newcommand{\efig}{\end{figure}}
\newcommand{\bcen}{\begin{center}}
\newcommand{\ecen}{\end{center}}

\long\def\comment#1{}

\def \n2{{N_0 \over 2}}

\def \h5{\hspace{0.5in}}

\newcommand{\bZ}{{\utwi{Z}}}

\newcommand{\dff}{\stackrel{\triangle}{=}}

\renewcommand{\baselinestretch}{1}

\def\IR{\mathbb R}

\renewcommand{\baselinestretch}{1.6}

\title{Statistical Non-linear Model, Achievable Rates and Signal Detection for Photon-level Photomultiplier Receiver}
\author{Zhimeng Jiang, Chen Gong, and Zhengyuan Xu
		\thanks{This work was supported by Key Program of National Natural Science Foundation of China (Grant No. 61631018) and Key Research Program of Frontier Sciences of CAS (Grant No. QYZDY-SSW-JSC003).}
		\thanks{The authors are with Key Laboratory of Wireless-Optical Communications, Chinese Academy of Sciences, University of Science and Technology of China, Hefei, Anhui 230027, China. 
			Email: zhimengj@mail.ustc.edu.cn, \{cgong821, xuzy\}@ustc.edu.cn.}}
\date{}
\begin{document}
\maketitle{}

\renewcommand{\baselinestretch}{1.3}

\begin{abstract}
	We characterize practical optical signal receiver in a wide range of signal intensity for optical wireless communication, from discrete pulse regime to continuous waveform regime. We first propose a statistical non-linear model based on the photomultiplier tube (PMT) multi-stage amplification and Poisson channel, and then derive the optimal and tractable suboptimal duty cycle with peak-power and average-power constraints for on-off key (OOK) modulation in the linear regime. Subsequently, a threshold-based classifier is proposed to distinguish the PMT working regimes based on the non-linear model. Moreover, we derive the approximate performance of mean power detection with ``infinite'' sampling rate and finite over-sampling rate in the linear regime based on short dead time assumption and central-limit theorem. We also formulate the performance from the perspective of communications in the non-linear regime.  Furthermore, the performance of mean power detection and photon counting detection under maximum likelihood (ML) criterion for different sampling rates is evaluated from both theoretical and numerical perspectives. We can conclude that the sample interval equivalent to dead time is a good choice, and lower sampling rate would significantly degrade the performance.   
\end{abstract}

%\newpage
\renewcommand{\baselinestretch}{1.5}

\section{Introductions} \label{sec.Introduction}
Optical wireless communication can work in both line-of-sight and non-line-of-sight (NLOS) scenarios \cite{xu2008ultraviolet}. On some specific occasions where conventional RF is
prohibited and direct-link transmission cannot be guaranteed,
NLOS ultra-violet optical scattering communication
provides an alternative solution to achieve certain information
transmission rate, but shows a large path loss. Optical wireless communication may work in a wide range of signal intensity, from continuous waveform regime to discrete pulse regime. For the latter, it is difficult to detect the
received signals using a conventional continuous waveform
receiver, such as photon-diode (PD) and avalanche photondiode
(APD). Instead, a photon-level photomultiplier tube (PMT) receiver needs to be employed, which converts the  received photons
to electronic pulse signals through multi-stage amplification.

Using a photon-level receiver, the number of detected photoelectrons satisfies a Poisson distribution, which forms a Poisson channel. For
Poisson channel, existing works mainly focus on the channel
capacity, such as the continuous Poisson
channel capacity \cite{wyner1988capacity}, discrete Poisson channel capacity \cite{lapidoth2009capacity}, wiretap Poisson channel capacity \cite{laourine2012degraded}, as well as the Poisson interference channel capacity \cite{lai2015capacity}. System characterization and optimization for binary inputs were investigated in \cite{el2012binary}. A variety of
channel estimation approaches have been proposed for indoor
visible light communication in \cite{hashemi2008channel,wu2012channel} and photon-counting PMT receiver in \cite{gong2016optical}.

Most information theory and signal process works focus on perfect photo-counting receiver. \cite{sarbazi2018statistical} investigates the counting statistics of active quenching
and passive quenching single photon avalanche diode (SPAD)
detectors. SPAD is only adopted to detect low signal intensity due to strong avalanche triggered a single electron-hole pair \cite{renker2009advances,zappa2002fully}, while PMT can detect wide range of signal intensity including photon-counting level signal and continuous waveform level signal \cite{liu2016signal}. %Thus, the characteristic of PMT is stringent to investigate. 
The characteristics of PMT, including single photoelectron spectrum, time properties, linearity and so on, are investigated in experiment \cite{xia2015performance,chirikov2001method,tripathi2003systematic,vasile1999high,musienko2006tests}. However, the model considering interior characteristic including single photoelectron spectrum and linearity is devoid.
A practical receiver typically consists of a PMT and the subsequent processing blocks \cite{becker2005advanced}. A practical solution is to adopt PMT to detect the arrival photons and generate a series of pulses with certain width, which incurs certain dead time. The receiver parameters optimization including pulse holding time and threshold has been investigated in \cite{zou2017characterization}, which shows negligible thermal noise compared with shot noise in experiments. Practical PMT signal characterization based on the three regimes and the non-linearity effect has been investigated in \cite{liu2016signal}. \cite{tubes2006basics} shows that the non-linear takes place between the last dynode and anode due to space charge effect. A novel PMT shot noise model with asymmetric probability density has been investigated in \cite{tan1982statistical} and outperforms Gaussian model reported by experimental PMT gain data \cite{delaney1964single}.

In this work, we aim to investigate the signal characterization, achievable transmission rate, and signal detection for a wide range of signal intensity, including discrete pulse regime, continuous waveform regime, and transition regime. More specifically, we propose a practical PMT model with a finite sampling rate analog-digital converter (ADC). We assume no electical thermal noise since it is negligible compared with shot noise in PMT and propose a statistical non-linear model on the PMT receiver. Based on the asymmetric shot noise model, we investigate the achievable rate at single and multiple sampling rates in the PMT linear regime, along with the optimal duty cycle and tractable suboptimal duty cycle. We propose a threshold-based classifier related to non-linear function to distinguish the discrete pulse regime, the continuous waveform regime, and the transition regime. Futhermore, we consider on-off keying (OOK) modulation and investigate the error probability of mean power detection (MPD) and photon counting detection (PCD) with different sampling rates. We derive approximate performance of the mean power detection with ``infinite" sampling rate and different over-sampling rates in the linear regime in the case of dead time shorter than the symbol duration and central-limit theorem. The theoretical results on the detection error probability are also evaluated and compared with simulation results.

The remainder of this paper is organized as follows. In Section~\ref{sec.System}, we present the PMT statistical non-linear model under consideration. In Section~\ref{sec.achierate}, we derive the optimal duty cycle and tractable suboptimal duty cycle for single- and multiple-sampling rates. In Section~\ref{sec.classPMT}, we propose a threshold-based classifier to distinguish the three work regimes. In Sections~\ref{sec.MPDinfi} and \ref{sec.detefi}, we investigate the error probability of MPD with ``infinite" and finite sampling rates.  Numerical results are given in Section~\ref{sec.NumericalResults}. Finally, Section~\ref{sec.Conclusion} provides the concluding remarks.

\section{System Model} \label{sec.System}
\subsection{PMT Principle Review}
\begin{figure}[htbp]
	\setlength{\abovecaptionskip}{0cm} %缩小caption和图像之间的距离
	\setlength{\belowcaptionskip}{0cm}
	\centering
	{\includegraphics[width=1\columnwidth]{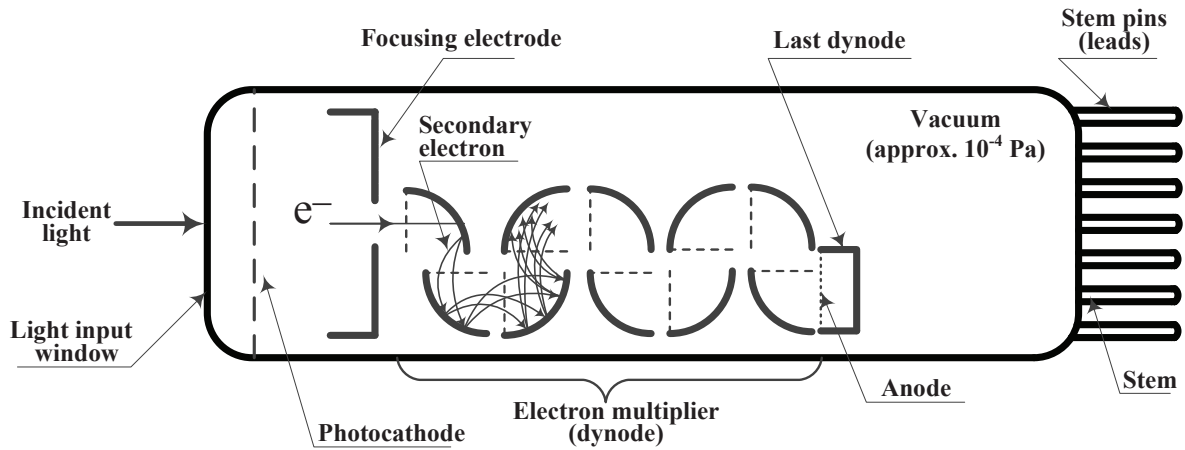}}
	\caption{The typical structure of a PMT.}
	\label{fig.pmtcons}
\end{figure}
The typical structure of a PMT is shown in Fig. \ref{fig.pmtcons}, including a photocathode, a focusing electrode, multiple dynodes (electron multiplier) and an anode. A single photon entering and detected by a PMT produces output signal through the following process. A single photon passes through the input window and excites electrons in the photocathode that are emitted into the vacuum. Such electrons are accelerated and focused by the focusing electrode onto the first dynode, where they are multiplied by means of secondary electron emission. Such secondary emission is repeated at each successive dynodes, and the multiplied secondary electrons emitted from the last dynode are finally collected by the anode.

When one or more photons arrive at the surface, the electrons in the valence band adsorb the photon energy and become excited, which are emitted into the vacuum if the diffused electrons have enough energy. The ratio of the output electrons over the incident photons is defined as the quantum efficiency. The emitted photoelectrons from the photocathode are focused onto the first dynode with certain collection efficiency, defined as the ratio of the number of electrons landing on effective area of the first dynode over the number of emitted photoelectrons. In this model, we merge the quantum efficiency and collection efficiency into the compound channel gain. The photoelectrons are accelerated in multiple dynodes with supply voltage from the voltage-divider circuits.

The PMT exhibits good linearity for the anode output current over a wide range of incident light power as well as in photon counting region. However, for too large incident light intensity, the output signal exhibits non-linearity characteristics for limited linearity of the cathode (secondary) and anode (primary), especially at a low supply voltage and large current. The upper limit of cathode linearity (average current) ranges from $0.01\mu A$ to $10\mu A$ depending on the photocathode materials \cite{tubes2006basics}. The anode linearity in the DC mode operation is primarily limited by the voltage-divider circuit, while that in the pulse mode operation is primarily limited by space charge effects, called current saturation phenomenon for large space charge density. In this model, we assume that the cathode is linear due to negligible current prior to amplification.

\subsection{PMT Gain for Single Photon}
Note that the number of secondary electrons excited per primary electron is Poisson distributed for small physical non-uniformities across the dynode surfaces \cite{chirikov2001method,prescott1966statistical}. Each electron is amplified by a multi-stage dynode, where the random
	electron multiplication process can be modelled by a Galton-Watson branching process. Defining $S_{\theta}$ as the total number of electrons emitted by the $\theta$-th dynode, we have $S_{\theta}=\sum_{i=1}^{S_{{\theta}-1}}N_{{\theta}i}$, where $\{N_{{\theta}i},{\theta}\geq1,i\geq1\}$ are independent Poisson random variables each with identical mean given by $\bar{h}$. The probability-generating function $\mathbb{E}[\omega^{S_{\nu}}]$ of $S_{\nu}$ is given by $m^{(\nu)}(\omega)$, where $\nu$ denotes the number of dynodes stages, $m^{(\theta)}(\omega)=m\big(m^{(\theta-1)}(\omega)\big)$ and $m^{(1)}(\omega)=e^{\bar{h}(\omega-1)}$. Define $G$ as the PMT gain for a single photon after $\nu$ stages. As its probability distribution appears to be intractable, \cite{tan1982statistical} adopts Markov diffusion process approximation to obtain the following moment-generating function (MGF),
	\be\label{eq.MGFG}
	M_G(\omega)=\exp(-\frac{{A\omega}}{1+B\omega}),
	\ee
	where $A=\mathbb{E}[G]$, $B=\frac{1}{2}\frac{A-1}{\sqrt[\nu]{A}-1}$, variance $\mathbb{D}[G]=2AB$ and $M_G(\omega)\dff\mathbb{E}[e^{-\omega G}]$.
\subsection{PMT Output Signal Formulation}
\begin{figure}[htbp]
	\setlength{\abovecaptionskip}{0cm} %缩小caption和图像之间的距离
	\setlength{\belowcaptionskip}{0cm}
	\centering
	{\includegraphics[width=1\columnwidth]{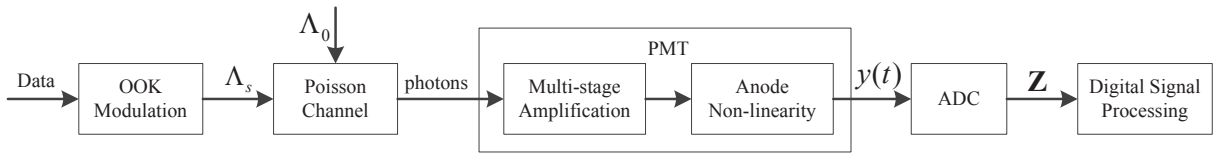}}
	\caption{The system model under consideration.}
	\label{fig.system}
\end{figure}
Consider a practical receiver with finite sampling rate, which contains a PMT detector, an ADC, and a digital signal processing unit. Upon a photon arriving, the PMT detector generates a
continuous pulse with certain width resulting from multi-stage amplification in the receiver dynode. The PMT output signals are sampled by the ADC, where the ADC outputs are employed for further digital signal processing. The
system model is shown in Fig. \ref{fig.system}. 

The transmitted signal denoted as $\Lambda_s$ satisfies the following peak-power and average power constraints, i.e., $0\leq\Lambda_s\leq\Lambda_{A}$ and $\mathbb{E}[\Lambda]\leq\eta\Lambda_{A}$, where $\Lambda_{A}$ and $\eta\Lambda_{A}$ denote the maximum peak power and maximum average power. Define $\xi \in \{0, 1\}$ as the OOK modulation symbol, corresponding to photon arrival rates $\Lambda_s\in\{0,\Lambda_{A}\}$. Defining $\Lambda_0$ as the background radiation intensity, the photon arrival time follows a Poisson process with certain arrival rate, denoted as $\Lambda$ given by $\Lambda=\Lambda_s+\Lambda_0$.
Defining $T_b$ and $N$ as the symbol duration and the number of detected photoelectrons during a symbol duration, respectively, we have
$\mathbb{P}(N=n)=e^{-\Lambda T_b}\frac{(\Lambda T_b)^n}{n!}\dff Poi(\Lambda T_b)$.

A single photon is amplified with a short time response due to different electron trajectories. Assume fixed time response duration causing dead time effect since the variation of hundred of picoseconds is negligible compared with time response duration about $20ns$ \cite{xia2015performance}. Define $\tau$ as dead time. Let $h(t)$ denotes the normalized single photoelectron spectrum, which is amplified by random PMT amplification gain. Defining $Y_j(t)$ as the photon impulse response in symbol duration $(-jT_b, (-j+1)T_b]$, we have
\be
\mathit{Y}_j(t)=\sum_{k=1}^{N_j}G^j_k h(t-t^j_k),
\ee
where $N_j$ denotes the number of arrived photons in $(-jT_b, (-j+1)T_b]$ with arrival time $t_k^j$ for $1 \leq k \leq N_j$, and the corresponding random PMT gains $G^j_k$ are independent and identically distributed satisfying the distribution given by Equation (\ref{eq.MGFG}).
For a wide range of signal intensity, the PMT saturation is caused by space charge effects or inadequate supply voltage between the last dynode and anode, and can be modeled by a non-linear function $C(\cdot)$. Hence, the output analog signal $y(t)$ can be written as follows,
\be 
y(t)=C\Big(\sum_{j=0}^{+\infty}\mathit{Y}_j(t)\Big).
\ee 
\iffalse
Denote ${\bf z}^{L}_1\dff\{z_1, z_2,\cdots, z_{L}\}$ as the samples from ADC, where $L$ denotes the number of samples. Here we omit the superscript, symbol duration index, since the arrival time within one symbol duration is of interest.
\fi

Consider the linear regime assuming $C(x)=x$ and $h(t)=u(t)-u(t-\tau)$, where $u(t)$ is the step function. In fact, rectangular $h(\cdot)$ can be achieved by pulse-holding circuits \cite{zou2017characterization,kruiskamp1994cmos,fletcher1975peak}.  Define $Z_j$ as the ADC $j$-th sampling value and $\tilde{Z}_j$ as the corresponding normalized sample value by average photon amplification $A$. For the photon rate $\Lambda$ and according to \cite{tan1982statistical}, we have the following MGF of $\tilde{Z}_j$ given photon arrival rate $\Lambda$,
\be\label{eq.MGFnorg}
M_{\tilde{Z}_j|\Lambda}(\omega)=e^{\lambda\big(\exp(-\frac{\omega}{1+a^{-1}\omega})-1\big)},
\ee 
where $\lambda=\Lambda\tau$ and $a=\frac{A}{B}$. Furthermore, the corresponding probability density function is given by
\be\label{eq.mixpdf}
f_{\tilde{Z}_j|\Lambda}(z)=e^{\lambda(e^{-a}-1)}\delta(z)+a\sqrt{\frac{1}{z}}e^{-(\lambda+az)}\sum_{n=0}^{\infty}\frac{\sqrt{n}(\lambda e^{-a})^n}{n!}I_1(2a\sqrt{nz}),
\ee 
where $I_1(\cdot)$ is the modified Bessel function of the first kind.

The important notations throughout this paper are summarized into Table~\ref{tab.Table1}.
\begin{table}
	\caption{Important notations adopted in this paper.}\label{tab.Table1}
	\centering
	\begin{tabular}{|p{2.5cm}|p{12cm}|}
		\hline
		Notation & Descriptions of Notation \\
		\hline
		$\Lambda_{s},\Lambda_{0}$ & The intensity of the transmitted signal, background \\
		\hline
		$\Lambda_{A},\xi$ & Peak power constraint and transmitted OOK symbol \\
		\hline
		$\lambda_{1},\lambda_{0}$ & The mean number of photons of the transmitted signal $\xi=1$ and $\xi=0$ in dead time \\
		\hline
		$G,M_G(\omega),f_G(z)$ & The PMT gain variable and corresponding MGF, probability density function \\
		\hline
		$T_s,T_b,\tau$ & Sample interval, symbol duration and dead time \\
		\hline
		$t_k^j,G_k^j$ & The arrival time of the $k^{th}$ photon in previous $j^{th}$ symbol duration and corresponding PMT gain\\
		\hline
		$N,\mathbf{T}^N$ & The number of arrival photons and corresponding arrival time sequence\\
		\hline
		$Y(t),y(t)$ & The output signal of last dynode and anode \\
		\hline
		$Z,\tilde{Z}$ & The samples from ADC and corresponding normalized samples \\
		\hline
		$I^C(\xi;\tilde{Z}),I^D(\xi;\tilde{Z})$ & The mutual information of continuous and discrete part $I(\xi;\tilde{Z})$ \\
		\hline
		$\mu_a^{*},\mu^{*}_1,\mu^{*}_{ma},\mu^{*}_{m}$ & The suboptimal and optimal suty cycle for single symbol-rate sampling and $m$ symbol-sampling rate \\
		\hline
		$\Lambda^{(1)}_{th},\Lambda^{(2)}_{th}$ & The proposed two threshold between three working regime \\
		\hline
		$r_j,y_s,M_{r_j},M_{y_s}$ & The mean power of previous photons response arrived in $j^{th}$ duty cycle and all photons response for infinite sampling rate, and corresponding MGF   \\
		\hline
		$k_s,k_d$ & The number of samples in dead time and the ratio of duty cycle over dead time for over-sampling rate \\
		\hline
		$y_s^{k_s},y_c^{k_s},y_l^{k_s}$ & The mean power of sum signal, current signal component and last symbol signal component for over-sampling rate \\
		\hline
		$\gamma_{MPD}^{La},\gamma_2^{*},\hat{n}_{th}$ & The Gaussian approximation decision threshold for MPD, the decision and counting threshold for PCD with under-sampling rate\\
		\hline
		$p_0,p_1,p_{th}$ & The probability of exceed the decision threshold for PCD given $\xi=0$, $\xi=1$, and counting decision related \\
		\hline
	\end{tabular}
\end{table}
In the remainder of this paper, we consider positive dead time and finite/infinite sampling rate at the receiver. We call under-sampling if the sampling interval is longer than the dead time and over-sampling otherwise.
\section{The achievable rate for finite sampling rate}\label{sec.achierate}
To analyze the achievable rate, we restrict the analysis on a specific symbol interval, and remove notation $j$ in this section for the arrival time and pulse amplification gains. Based on Section \uppercase\expandafter{\romannumeral2}, given the number of arrived photons $N$, we define $\mathbf{T}^N \dff [t_1, t_2, ..., t_N]$ as the corresponding arrival time. Then, the received signal $y(t)$ is determined by pair $(N,\mathbf{T}^{N})$, as well as the normalized single photoelectron spectrum $h(\cdot)$. Assume duty cycle $\mu$ for the OOK modulation. Let $T_s$ denote the sampling interval and $\bZ$ denote the samples within a symbol interval. Note that $y(\cdot)$ is one-to-one with $(N,\mathbf{T}^{N})$ if $C(x)=x$ and constant PMT gain $G_{k}$, which implies mutual information $I(\xi;(N,\mathbf{T}^{N}))=I(\xi;y(t))$. Since $\xi\longrightarrow (N,\mathbf{T}^{N}) \longrightarrow y \longrightarrow {\bZ}\longrightarrow \hat{N}$ forms a Markov chain, we have the general expression $I(\xi;(N,\mathbf{T}^{N}))\geq I(\xi;y(t))\geq I(\xi;\bZ)$, where strictly larger sign typically holds due to the shot noise of amplification gain and finite sampling rate.

In this section, we investigate the achievable rate with finite sampling rate in the linear regime with $C(x)=x$ assuming $h(t)=u(t)-u(t-\tau)$. Note that $y(t)=\sum_{k=1}^{N}G_{k}\mathbbm{1}\{t_k-t+\tau>0\}\mathbbm{1}\{t-t_k>0\}$, where $\mathbbm{1}\{\cdot\}$ is the indicator function. Defining $T_s$ as the sampling interval,  the output samples are i.i.d. if $T_s>\tau$. We consider both single sample and multiple samples in a symbol duration, called single symbol-rate sampling and multiple symbols-rate sampling. 
\iffalse
We have ADC samples $Z_i=\sum_{n=1}^{N_1}G_{in}$, where the number of photons $N_1$ arriving at $(iT_s-\tau,iT_s]$ satisfies $N_1\sim Poi(\Lambda\tau)$. Defining $\tilde{Z}=\frac{Z}{A}$ and $\mu$ as the duty cycle, for the number of samples $n_1>n_2$ and optimal duty cycle $\mu^*$ for $n_2$ samples, we have
\be 
\max\limits_{0\leq\mu\leq\eta}I(b;\tilde{\bf z}^{n_2}_{1})=I(b;\tilde{\bf z}^{n_2}_{1}|_{\mu=\mu^{*}})&\leq& I(b;\tilde{\bf z}^{n_1}_{1}|_{\mu=\mu^{*}})\\ \nonumber &\leq& \max\limits_{0\leq\mu\leq\eta}I(b;\tilde{\bf z}^{n_1}_{1}),
\ee
which shows that the maximum achievable rate increases with the number of samples in a symbol duration.
\fi

\subsection{Single Symbol-rate Sampling}
According to the above model, we have the conditional probability function for symbols zero and one as follows:
\be 
f_{\tilde{z}|\xi=i}=f_{\tilde{Z}|\lambda_i}(\tilde{z}), i=0,1,
\ee 
where $\lambda_0=\Lambda_0\tau$, $\lambda_1=\Lambda_1\tau$ and $\Lambda_1=\Lambda_0+\Lambda_A$. Let $\tilde z$ denote the normalized sample and $\tilde Z$ denote the corresponding random version. For sufficiently small $\Lambda_0$, we have the following result up to the first order of $\lambda_0$,
\be 
f_{\tilde{Z}|\lambda_0}(\tilde{z})&=&\sum_{i=0}^{+\infty}f_{\tilde{Z}}(\tilde{z}|N=i)\mathbb{P}(N=i|\lambda_0) \\ \nonumber
&=& e^{-\lambda_0}\delta(\tilde{z})+\lambda_0 e^{-\lambda_0}[e^{-a}\delta(\tilde{z})+ae^{-a(1+\tilde{z})}\sqrt{\frac{1}{\tilde{z}}}I_1(2a\sqrt{\tilde{z}})]+o(\lambda_0)\\ \nonumber
&=&(a_0+o(\lambda_0))\delta(\tilde{z})+\big(\lambda_0+o(\lambda_0)\big)g(\tilde{z}),
\ee 
where $g(\tilde{z})=ae^{-a(1+\tilde{z})}\sqrt{\frac{1}{\tilde{z}}}I_1(2a\sqrt{\tilde{z}})$ and $a_0=1-(1-e^{-a})\lambda_0$. Similarly to the scenario for $\lambda = \lambda_0$, based on Equation (\ref{eq.mixpdf}) we can obtain the probability density function for $\lambda = \lambda_1$. Note that probability function $f_{\tilde{Z}|\lambda}(\tilde{z})$ can be decomposed into discrete part $f_{\tilde{Z}|\lambda}^{D}(\tilde{z})$ and continuous part $f_{\tilde{Z}|\lambda}^{C}(\tilde{z})$, with the conditional cumulative distribution $F^D_{\tilde{Z}|\lambda}(\tilde{z})$ and $F^C_{\tilde{Z}|\lambda}(\tilde{z})$, respectively. 

To further characterize the mutual information, we first derive the uniform
continuity of $f_{\tilde{Z}|\lambda}^{C}(\tilde{z})$, as formalized by the following result.
\begin{lemma}\label{lem.1}
	Function $f_{\tilde{Z}|\lambda}^{C}(\tilde{z})$ is uniformly continuous
	with respect to $(\tilde{z},\lambda)\in\mathbf{R}_+\times(0,\lambda_A]$.
	\begin{proof}
		Please refer to Appendix \ref{appd.lem1}.
	\end{proof}	
\end{lemma}

Considering the distribution of $\tilde{Z}$ consisting of continuous and discrete parts, we have that the mutual information can be expressed as the following summation of the continuous part and discrete part, given by
\be\label{eq.mutuasingle}
I(\xi;\tilde{Z})=\int\log\frac{\mathrm{d}F_{\xi,\tilde{Z}}(\xi,\tilde{z})}{\mathrm{d}F_{\xi}(\xi)\mathrm{d}F_{\tilde{Z}}(\tilde{z})}\mathrm{d}F_{\xi,\tilde{Z}}(\xi,\tilde{z})\dff I^{D}(\xi;\tilde{Z})+I^{C}(\xi;\tilde{Z}),
\ee 
where the mutual information for the discrete part is given as follows,
\be\label{eq.Idis}
I^{D}(\xi;\tilde{Z})=-(\mu a_1+(1-\mu)a_0)\log(\mu a_1+(1-\mu)a_0)+[\mu a_1\log a_1 +(1-\mu)a_0\log a_0],
\ee
where $a_i=e^{-\lambda_i(1-e^{-a})}$ for $i=0,1$. The mutual information for the continuous part is given as follows, 
\be\label{eq.Icon}
I^{C}(\xi;\tilde{Z})=H\big(\mu f_{\tilde{Z}|\lambda_1}^{C}(\tilde{z})+(1-\mu)f_{\tilde{Z}|\lambda_0}^{C}(\tilde{z})\big)-\mu H\big(f_{\tilde{Z}|\lambda_1}^{C}(\tilde{z})\big)-(1-\mu)H\big(f_{\tilde{Z}|\lambda_0}^{C}(\tilde{z})\big),
\ee 
where $H(f)\dff-\int f(\tilde{z})\log f(\tilde{z})\mathrm{d}\tilde{z}$. Define $H_2(f_1,f_2)\dff-\int f_1(\tilde{z})\log f_2(\tilde{z})\mathrm{d}\tilde{z}$ and thus $H(f)=H_2(f,f)$. Since brute-force calculation of $H(f_{\tilde{Z}|\lambda}^{C}(\tilde{z}))$ is intractable and $\lambda_0$ is negligible compared with $\lambda_s$, we resort to Taylor's expansion to approximate $I^{C}(\xi;\tilde{Z})$ under small $\lambda_0$.
To further characterize the mutual information approximation, we have the following Lemma on the relationship of $f_{\tilde{Z}|\lambda_0}^{C}$ and $f_{\tilde{Z}|\lambda_1}^{C}$.
\begin{lemma}\label{lemma.epsilon}
	For any fixed $\lambda_1$ and sufficiently small $\epsilon$ such that $0<\epsilon<e$, there exists $\delta$ independent of $\tilde{z}$ such that for any $0<\lambda_0<\delta$, 
	$f_{\tilde{Z}|\lambda_0}^{C}(\tilde{z})<\epsilon f_{\tilde{Z}|\lambda_1}^{C}(\tilde{z})$.
	\begin{proof}
		Please refer to Appendix \ref{appd.lem2}.
	\end{proof}	
\end{lemma}

Based on Lemma \ref{lemma.epsilon}, we have the following Taylor's expansion of $I^{C}(\xi;\tilde{Z})$,
\begin{lemma}\label{lemma.appmutual}
	The first-order Taylor's expansion of $I^{C}(\xi;\tilde{Z})$ is given as follows,
	\be 
	I^{C}(\xi;\tilde{Z})&=&-\mu\log\mu\big(1-(e^{-\lambda_1(1-e^{-a})})\big)-(1-\mu)\log\mu\lambda_0(1-e^{-a})\\ \nonumber
	&&-(1-\mu)\lambda_0(1-e^{-a})+\lambda_0(1-\mu)(C_1-C_2)+o(\lambda_0).
	\ee 
	where $C_1=-\int_{0}^{+\infty}g(\tilde{z})\log f_{\tilde{Z}|\lambda_1}^{C}(\tilde{z})\mathrm{d}\tilde{z}$ and $C_2=-\int_{0}^{+\infty}g(\tilde{z})\log g(\tilde{z})\mathrm{d}\tilde{z}$.
	\begin{proof}
		Please refer to Appendix \ref{appd.apprmutual}.
	\end{proof}	
\end{lemma}

Noting that $\log(\mu a_1+(1-\mu)a_0)=\log(\mu a_1+1-\mu)-\frac{(1-\mu)(1-e^{-a})}{\mu a_1+1-\mu}\lambda_0+o(\lambda_0)$, we have
\be 
I^{D}(\xi;\tilde{Z})&=&-(\mu a_1+1-\mu)\log(\mu a_1+1-\mu)+\mu a_1\log a_1\\ \nonumber
&&+(1-\mu)(1-e^{-a})\log(\mu a_1+1-\mu)\lambda_0+o(\lambda_0),
\ee 
and thus mutual information $I(\xi;\tilde{Z})$ can be given as follows,
\be
I(\xi;\tilde{Z})&=&-\mu\log\mu(1-a_1)-(\mu a_1+1-\mu)\log(\mu a_1+1-\mu)+\mu a_1\log a_1\\ \nonumber
&&+(1-\mu)(1-a_0)\log(a_1-1+\frac{1}{\mu})+\lambda_0(1-\mu)(C_1-C_2)+o(\lambda_0).
\ee
Due to small $\lambda_0$ in practical scenarios, we omit the terms that attenuate with $\lambda_0$ to obtain a suboptimal duty cycle $\mu_a^{*}$. The suboptimal duty cycle $\mu_a^{*}$ and optimal duty cycle $\mu_1^{*}$ are defined as follows,
\be\label{eq.probsub}
\mu_a^{*}&\dff&\arg\max\limits_{0\leq\mu\leq\eta}\{-\mu\log\mu(1-a_1)-(\mu a_1+1-\mu)\log(\mu a_1+1-\mu)+\mu a_1\log a_1\},\\
\mu_1^{*}&\dff&\arg\max\limits_{0\leq\mu\leq\eta}I(\xi;\tilde{Z}).
\ee 
\begin{theorem}\label{theo.submu1}
	For $0<a_1<1$, the suboptimal duty cycle $\mu^{*}_a$ is given by $\mu_a^{*}=\min\{\eta,\mu_a\}$, where $\mu_a=(1-a_1+a_1^{-\frac{a_1}{1-a_1}})^{-1}$. Moreover, we have $\lim\limits_{\lambda_A\to0}\mu_a=\frac{1}{e}$, $\lim\limits_{\lambda_A\to+\infty}\mu_a=\frac{1}{2}$.
	\begin{proof}
	Please refer to Appendix \ref{appd.theor1}.		
	\end{proof}
\end{theorem}

Note that the above asymptotic result is different from that reported in \cite{wyner1988capacity} where for $\lambda_0 = 0$ the optimal duty cycle is $e^{-1}$. Note that in this work, since random PMT amplification is incorporated, the asymptotic optimal duty cycle varies changes to $1/2$ as $\lambda_A$ approaches infinity. Moreover, we have the following on the optimal duty cycle.
\begin{theorem}\label{theor.optmu1}
	For $0<a_1<1$, the optimal duty cycle $\mu_1^{*}=\min\{\eta,\mu_1\}$, where $\mu_1$ is the unique solution to the following equation
	\be\label{eq.suboptmu}
	&&\int_{0}^{+\infty}\big(f_{\tilde{Z}|\lambda_0}^{C}(\tilde{z})-  f_{\tilde{Z}|\lambda_1}^{C}(\tilde{z})\big)\log[\mu f_{\tilde{Z}|\lambda_1}^{C}(\tilde{z})+(1-\mu)f_{\tilde{Z}|\lambda_0}^{C}(\tilde{z})]\mathrm{d}\tilde{z}-H(f_{\tilde{Z}|\lambda_1}^{C})\\ \nonumber &&+H(f_{\tilde{Z}|\lambda_0}^{C})-(a_1-a_0)\log[\mu a_1+(1-\mu)a_0]+a_1\log a_1-a_0\log a_0=0.
	\ee
	\begin{proof}
		Please refer to Appendix \ref{appd.theor2}.
	\end{proof}
\end{theorem}
\subsection{Multiple Symbols-rate Sampling}
For simplicity, we use the following notations. Let $\Omega\dff\{1,2,\cdots,L\}$, where $L$ denotes the number of samples in a symbol duration. Define set $\tilde{Z}^S\dff \{Z_i|i\in S\}$ for $S\subseteq\Omega$, set $S^C\dff\Omega-S$, probability density function $f_{S|j}\dff f_{\tilde{z}^S|\xi=j}$, product $\delta(\tilde{z}^S)\dff\prod\limits_{i\in S}\delta(\tilde{z}_i)$, and $f_{S|j}^C$ as the continuous part of $f_{S|j}$.

We have the following results on $I(\xi;\tilde{Z}^\Omega)$.
\begin{lemma}\label{lem.mutumulti}
	Assume $T_s\geq\tau$, then mutual information $I(\xi;\tilde{Z}^\Omega)$ is the summation of the following $2^{L}$ items,
	\be 
	I(\xi;\tilde{Z}^\Omega)=\sum\limits_{S\subseteq\Omega}I^S(\xi,\tilde{Z}),
	\ee 
	where $I^S(\xi,\tilde{Z})=H(\mu a_1^{|S|}f_{S^C|1}^{C}+(1-\mu)a_0^{|S|}f_{S^C|0}^{C})-\mu H(a_1^{|S|}f_{S^C|1}^{C})-(1-\mu)H(a_0^{|S|}f_{S^C|0}^{C})$, $f_{\o|j}^{C}=1$ and $H(a_j^{|\Omega|}f_{\o|j}^{C})=-a_j^{|\Omega|}\log(a_j^{|\Omega|})$.
	\begin{proof}
		Please refer to Appendix \ref{appd.mutumulti}.
	\end{proof}	
\end{lemma}

The suboptimal duty cycle $\mu_{ma}^{*}$ and optimal duty cycle $\mu_m^{*}$ are defined as follows:
\be 
\mu_{ma}^{*}&=&\arg\max\limits_{0\leq\mu\leq\eta}I(\xi;\tilde{Z}^\Omega|_{\lambda_0 = 0}),\\
\mu_m^{*}&=&\arg\max\limits_{0\leq\mu\leq\eta}I(\xi;\tilde{Z}^\Omega).
\ee
The characteristics of suboptimal duty cycle $\mu_{ma}^{*}$ and optimal duty cycle $\mu_m^{*}$ are shown as follows,
\begin{theorem}\label{theor.3}
	For multiple samples, the optimal duty cycle $\mu_m^{*}=\min\{\eta,\mu_m\}$, where $\mu_m$ is the unique solution to the following,
	\be\label{eq.solumulti}
	&&\sum\limits_{S\subseteq\Omega}\int_{\mathbf{R}_{+}^{|\Omega|-|S|}}\big(a_0^{|S|}f_{S^C|0}^C(\tilde{z}^{S^C})-a_1^{|S|}f_{S^C|1}^C(\tilde{z}^{S^C})\big)\log\big(\mu a_1^{|S|}f_{S^C|1}^{C}(\tilde{z}^{S^C})\\ \nonumber
	&&+(1-\mu)a_0^{|S|}f_{S^C|0}^{C}(\tilde{z}^{S^C})\big)\mathrm{d}\tilde{z}^{S^C}-H(a_1^{|S|}f_{S^C|1}^C)+H(a_0^{|S|}f_{S^C|0}^C)=0.
	\ee
	\begin{proof}
	Please refer to Appendix \ref{appd.theor3}.
	\end{proof}
\end{theorem} 

Moreover, we have the following approximated solution under small background radiation $\lambda_0$, similarly to that for single symbol-rate sampling.
\begin{theorem}\label{theor.4}
	For multiple symbols-rate samples, a suboptimal duty cycle based on small $\lambda_0$ is given by $\mu_{ma}^{*}=\min\{\eta,\mu_{ma}\}$, where $\mu_{ma}=(1-a_1^{L}+(a_1^{L})^{-\frac{a_1^{L}}{1-a_1^{L}}})^{-1}$.
	\begin{proof}
 	 Please refer to Appendix \ref{appd.theor4}.
	\end{proof}
\end{theorem}
\begin{figure}[htbp]
	\setlength{\abovecaptionskip}{-0.2cm} %缩小caption和图像之间的距离
	\setlength{\belowcaptionskip}{-0.2cm}
	\centering
	{\includegraphics[width =4.5in]{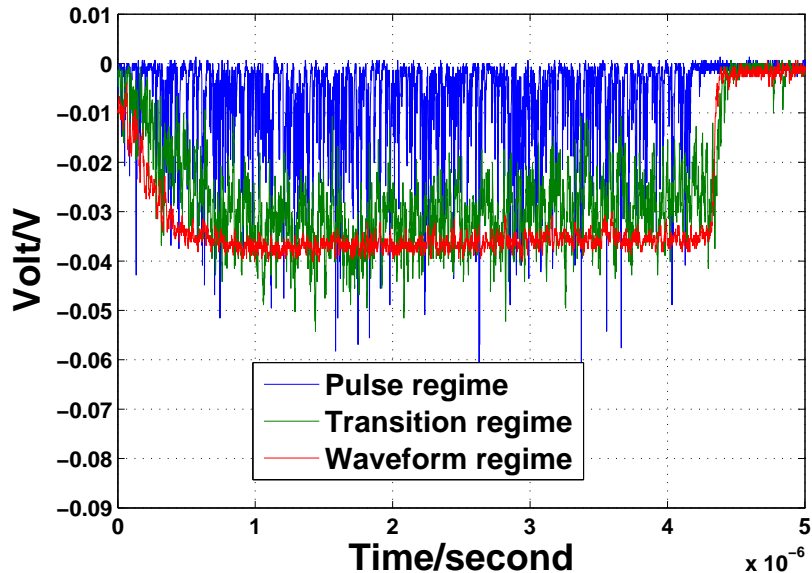}}
	\caption{The typical signals of three PMT working regimes from experiment.}
	\label{fig.threesignal}
\end{figure}

\section{working regime classification of PMT} \label{sec.classPMT}
In general, the PMT signal range can be divided into three regimes,  photon counting regime, transition regime and waveform regime, according to the received signal intensity, where the experimental results are shown in Fig. \ref{fig.threesignal}. In the experiment, we employ RIGOL DG5252 AWG to transmit random data to a blue LED, Hamamatsu CR315 PMT $465nm$ to detect the
optical signal and  Agilent MSOX6004A oscilloscope to collect the PMT output signal. The high level voltages for LED of the pulse regime, the transition regime and the waveform regime are $2.30V$, $2.36V$, $2.42V$, respectively. A fundamental question is how to analytically characterize the three regimes. In this section, the symbol duration is normalized such that $T_b=1$ and we propose a threshold-based criterion. Note that the pulses are generated by a practical PMT detector with fixed holding time, which enables the pulse detection via finite-rate sampling. However, a photon detected at time $t$ creates a holding time interval from $t$ to $t+\tau$, during which the photon arrival cannot be recorded accurately.

Assuming infinite sampling rate and zero noise variance of the PMT detector, the probability mass function (PMF) of the detected photons number $n$ based on rising-edge detection can be characterized by the following result \cite{omote1990dead}. Note that in such a scenario, the photons arrived within duration $\tau$ of an arrived photon cannot be detected, which accords with the basic assumption of work \cite{omote1990dead}.
\begin{Proposition}
	Given normalized dead time $\tau$ and photon arrival rate $\Lambda$, the number of detected pulses $n$ under infinite rate sampling is characterized by the following probability function,
	\be\label{eq.deadcount} 
	\mathbb{P}(n|\Lambda,\tau)=\sum_{m=0}^{M-n}\frac{(-1)^m}{n!m!}\big[\big(1-(n+m-1)\tau\big)\Lambda e^{-\Lambda\tau}\big]^{n+m},
	\ee  
	where interger $M\dff\lfloor\frac{1}{\tau}\rfloor+1$ defines the maximum number of counted pulses. Moreover, the mean and variance of $n$ are given as follows
	\be
	\mathbb{E}[n]&=&\Lambda e^{-\Lambda\tau},\\
	\mathbb{D}[n]&=&\mathbb{E}[n]-[1-(1-\tau)^2]\mathbb{E}[n]^2.
	\ee 
\end{Proposition}

Let $\Lambda^{(1)}_{th}$ denote the threshold between the photon counting regime and transition regime, and $\Lambda^{(2)}_{th}$ denote the threshold between the transition regime and waveform regime. We propose the following threshold specification criterion.

\textbf{Threshold Between Pulse Regime and Transition Regime}: Let $\Lambda^{(1)}_{th}=\arg\max\limits_{\lambda}\mathbb{E}[N]$, where random variable $N$ follows the probability mass function given by Equation (\ref{eq.deadcount}).
Via simple calculation, we have that the threshold between the pulse regime and transition regime is given by $\Lambda^{(1)}_{th}=\frac{1}{\tau}$.

\textbf{Threshold Between Transition Regime and Waveform Regime}: Consider function $C$ with domain $[0,+\infty)$, which is bounded and smooth, satisfying $C(0)=0$, $C^{'}(0^{+})>0$ and $\lim\limits_{x\to+\infty}C(x)$ exists. Consider $l_{max}\dff\sup\limits_{\mu_L(\{x:C(x)\leq l\})<+\infty} l$, where $\mu_L(\cdot)$ denotes Lebesgue measure. The waveform regime is defined according to the probability of output samples lower than $l_{max}$. More specifically, given probability threshold $\epsilon$, the threshold between the transition regime and waveform regime $\Lambda^{(2)}_{th}$ is given by the maximum $\Lambda_d$ where such probability is lower than or equal to $\epsilon$, i.e.,
\begin{equation}\label{eq.optgam2}
\begin{aligned}
& \max
& &\Lambda_d,\\
& \text{s.t.}
& &\mathbb{P}(\tilde{Z}\leq l_{max}|\Lambda)\leq\epsilon, \forall\Lambda\geq\Lambda_d.
\end{aligned}
\end{equation}
\begin{lemma}\label{lemma.exist}
	The optimal solution to Problem (\ref{eq.optgam2}) exists and is unique for any $\epsilon>0$. 
	\begin{proof}
		Please refer to Appendix \ref{appd.lemexist}.				
	\end{proof}
\end{lemma}

Note that the optimal solution to Problem (\ref{eq.optgam2}) is intractable due to the complicated distribution of $\tilde{Z}$. We propose a suboptimal solution that can well approximate for large $\lambda$. Given $l_{max}$ and small $\epsilon$, we have $\Lambda^{(2)}_{th}>>l_{max}$, which implies the validity of adopting Gaussian approximation. More specifically, we have the following result.  
\begin{lemma}\label{lem.gauap}
	Define $\tilde{Z}_{nor}=\frac{\tilde{Z}-\mathbb{E}[\tilde{Z}]}{\sqrt{\mathbb{D}[\tilde{Z}]}}$ as the Gaussian normalization of $\tilde Z$. Then, as $\Lambda$ approaches infinity, $\tilde{Z}_{nor}$ converges in distribution to a standard normal distribution $\mathcal{N}(0,1)$.
	\begin{proof}
		  Please refer to Appendix \ref{appd.lemgauapp}.
	\end{proof}
\end{lemma}
\iffalse
\begin{remark}
	 MGF method is a more general method than Kullback–Leibler divergence which shows the asymptotic Gaussian distribution proterty. \cite{moser2012capacity} shows that the continuous Poisson random viable $T_c$, defined as $T_c\dff T+U$, where $T\sim\mathcal{P}oi(\lambda)$, $U\sim\mathcal{U}\Big([0,1)\Big)$ and $U$ is independent of $T$, will approximate a Gaussian distribution with mean $\lambda$ and variance $\lambda$ by Kullback–Leibler divergence. We give a new proof in view of MGF. 
	 
	 Noted that $M_T(\mu)=e^{\lambda(e^{-\mu}-1)}$ and $M_U(\mu)=\frac{1-e^{-\mu}}{\mu}$, thus $M_{T_c}(\mu)=e^{\lambda(e^{-\mu}-1)}\frac{1-e^{-\mu}}{\mu}$. Define normalization $S=\frac{T_c-\lambda}{\sqrt{\lambda}}$, then we have
	 \be 
	 M_{S}(\mu)=e^{\lambda(1-e^{\frac{-\mu}{\sqrt{\lambda}}})}\frac{1-e^{\frac{-\mu}{\sqrt{\lambda}}}}{\frac{\mu}{\sqrt{\lambda}}}e^{\lambda\mu}=e^{\frac{1}{2}\mu^2}+\frac{\mu}{2}e^{\frac{1}{2}\mu^2}\frac{1}{\sqrt{\lambda}}+o(\frac{1}{\sqrt{\lambda}})\overset{\lambda\to+\infty}{\longrightarrow}\mathcal{N}(0,1).
	 \ee 
\end{remark}
\fi

Based on the above results, we have the following on the Gaussian approximation-based threshold between the transition regime and waveform regime.
\begin{theorem}\label{theor.5}
	Given a small $\epsilon$, an approximate solution to Problem (\ref{eq.optgam2}) is given by
	\be 		\Lambda^{(2)}_{th}=\frac{1}{4\tau}\big(-\Phi^{-1}(\epsilon)\sqrt{1+2a^{-1}}+\sqrt{(1+2a^{-1})[\Phi^{-1}(\epsilon)]^2+4l_{max}}\big)^2,
	\ee
	where $\Phi(y)=\int_{-\infty}^{y}\frac{1}{\sqrt{2\pi}}e^{-\frac{x^2}{2}}\mathrm{d}x$.
	\begin{proof}
		Please refer to Appendix \ref{appd.theor5}.
	\end{proof}
\end{theorem}
\section{mean power detection For Over-sampling}\label{sec.MPDinfi}
Assume transmitted signal $\Lambda_s\in\{0,\Lambda_A\}$ with duty cycle $\mu=\frac{1}{2}$ and we investigate the binary hypothesis testing problem. For $\Lambda_s T_b\ll l_{max}$, the PMT is working in the linear regime $C(y)=y$. In this section, the performance of MPD with infinite sampling rate and sampling interval shorter than the dead time in the linear regime is investigated assuming symbol duration $T_b=1$ for simplicity.
\subsection{Infinite Sampling Rate in Linear Regime}
For signal $y(t)$ in symbol duration $[0,1]$, the mean power $y_s\dff\int_{0}^{1}y(t)\mathrm{d}t$, is used for symbol detection. Since each photon response is linear superimposed, each photon response can be calculated as follows. Defining the mean power of photon response with arrival time $t_k^j$ as $a_j(t_k^j)\dff\int_{0}^{1}s(t;t_k^j)\mathrm{d}t=G^j_k\alpha(t_k^j)$, where $\alpha(t_k^j)=\int_{0}^{1}h(t-t_k^j)\mathrm{d}t$, the mean power of response of photons arrived in symbol duration $[-j,-j+1)$ is given by $r_j\dff\sum_{k:-j\leq t_k^j<-j+1}a_j(t_k^j)$ and the mean power $y_s=\sum_{j=0}^{+\infty}r_j$.

Define $M_{r_0|\Lambda_s+\Lambda_0}$ and $M_{r_0|\Lambda_0}$ as the MGF of $r_0$ for symbol on and off, respective. Define $M_{y_s|\xi}$ as the MGF of $y_s$ given symbol $\xi$, and $M_{r_j}$ as the MGF of $r_j$ for $j \geq 0$. We have the following result.
\begin{lemma}\label{lemma.meanpower}
	Assume transmitted signal $\Lambda_s\in\{0,\Lambda_A\}$ with duty cycle $\mu=\frac{1}{2}$, the expression of $M_{y_s|\xi}(\omega)$ given symbol $\xi$ is as follow,
	\be\label{eq.MGF_Y}
	M_{y_s|\xi}(\omega)=\left\{\begin{array}{ll}
		M_{r_0|\Lambda_s+\Lambda_0}(\omega)\prod_{j=1}^{+\infty}M_{r_j}(\omega),&\xi=1;  \\
		M_{r_0|\Lambda_0}(\omega)\prod_{j=1}^{+\infty}M_{r_j}(\omega),&\xi=0.  
	\end{array}\right.
	\ee 
	\begin{proof}
		Please refer to Appendix \ref{appd.meanpower}.
	\end{proof}
\end{lemma}

Since $\Lambda_0\ll\Lambda_s$, we have $M_{r|\Lambda_0}(\omega)\approx 1$ and $M_{r_j|\Lambda_s+\Lambda_0}(\omega),j=1,2,\cdots$ is negligible, and thus
\be
\prod_{j=1}^{n}M_{r_j}(\omega)&=&\exp\Big(\sum_{j=1}^{n}\ln{M_{r_j}(\omega)}\Big)\approx \exp\Big(\frac{1}{2}\{\sum_{j=1}^{n}M_{r_j|\Lambda_s+\Lambda_0}(\omega)-1\}\Big)\\&=&\exp\Big(\frac{1}{2}\{\sum_{j=1}^{n}(\exp((\Lambda_s+\Lambda_0)[\int_{-j}^{-j+1}M_{a_j|\rho}(\omega)d\rho-1]))-1\}\Big)\\  \nonumber
&\approx& \exp\Big(\frac{(\Lambda_s+\Lambda_0)}{2}\sum_{j=1}^{n}[\int_{-j}^{-j+1}M_{a_j|\rho}(\omega)-1d\rho]\Big).%\nonumber
%&\longrightarrow& exp(\frac{(\Lambda_s+\Lambda_0)}{2}[\int_{-\infty}^{0}M_{a_j|\rho}(\omega)-1d\rho])
\ee
The approximation of $M_{y_s|\xi}(\omega)$ is obtained based on Lemma \ref{lemma.meanpower} as follows,
\be 
M_{y_s|\xi}(\omega)\approx\left\{\begin{array}{ll}
	\exp\Big(\Lambda_s^{'}\sum_{j=1}^{+\infty}[\int_{-j}^{-j+1}M_{a_j|\rho}(\omega)-1d\rho]+\Lambda_s\int_{0}^{1}M_{a_0|\rho}(\omega)-1\mathrm{d}\rho\Big),&\xi=1;  \\
	\exp\Big(\Lambda_s^{'}\sum_{j=1}^{+\infty}[\int_{-j}^{-j+1}M_{a_j|\rho}(\omega)-1d\rho]+\Lambda_0\int_{0}^{1}M_{a_0|\rho}(\omega)-1\mathrm{d}\rho\Big),&\xi=0; 
\end{array}\right.
\ee  
where $\Lambda_s^{'}=\frac{(\Lambda_s+\Lambda_0)}{2}$. Then, we have the following results on the conditional PDF $f_{y_s}(x|\xi)$ of $y_s$.

%\be 
%f_{y_s}(x|\xi)=\frac{1}{2\pi}\int_{-\infty}^{+\infty}M_{y_s|\xi}(if)e^{i2\pi ifx}\mathrm{d}f.
%\ee 
%According to the  first order condition, we have the best decesion threshold is solution about $s$ of following equation:
%\be 
%\int_{-\infty i}^{+\infty i} e^{(\Lambda_s+\Lambda_0)\int_{-\infty}^{0}M_{a_j}(\omega|\xi)-1\mathrm{d}\xi-\omega x}[e^{(\Lambda_s+\Lambda_0)\int_{0}^{1}M_{a_j}(\omega|\xi)-1d\xi}-1] \mathrm{d}\omega=0
%\ee 
\begin{corollary}\label{corol.meanpower2}
	For a photon response $h(t)=u(t)-u(t-\tau)$ and normalize $G$ by $\mathbb{E}[G]$, then we have
	\be 
	M_{y_s|\xi=1}(\omega)&=&\exp[-(\Lambda_s+\Lambda_0)(1-\tau)(1-e^{-\frac{\omega\tau}{1+a^{-1}\omega\tau}})]\exp[-\frac{3\Lambda_s^{'}}{2}\tau^2\omega]+o(\tau^2), \\
	M_{y_s|\xi=0}(\omega)&=&\exp[-\Lambda_0(1-\tau)(1-e^{-\frac{\omega\tau}{1+a^{-1}\omega\tau}})]\exp[-\frac{(\Lambda_s^{'}+\Lambda_0)}{2}\tau^2\omega]+o(\tau^2),
	\ee 
	where $\Lambda_s^{'}=\frac{\Lambda_0+\Lambda_s}{2}$.	
	\begin{proof}
		Please refer to Appendix \ref{appd.meanpower2}.
	\end{proof}	
\end{corollary}

According to Corollary \ref{corol.meanpower2} and omit $o(\tau^2)$ item, we have the following approximation on the distributions of $y_s$,
\be 
f_{y_s}(x|\xi=1)&\approx& \frac{1}{\tau}f_{\tilde{Z}}\Big(\frac{x}{\tau}-\frac{3\Lambda_s^{'}}{2}\tau^2|(\Lambda_s+\Lambda_0)(1-\tau)\Big),\\
f_{y_s}(x|\xi=1)&\approx& \frac{1}{\tau}f_{\tilde{Z}}\Big(\frac{x}{\tau}-\frac{(\Lambda_s^{'}+\Lambda_0)}{2}\tau^2|\Lambda_0(1-\tau)\Big).
\ee 

Consider the ML detection, where the regions corresponding to $\xi=0$ and $\xi=1$ are given as follows:
\be 
\delta_0(x)=\mathbbm{1}\{f_{y_s}(x|\xi=0)>f_{y_s}(x|\xi=1)\}, \\ 
\delta_1(x)=\mathbbm{1}\{f_{y_s}(x|\xi=0)\leq f_{y_s}(x|\xi=1)\}.
\ee 
Then, the detection error probability can be approximated as follows,
\be\label{eq.BERintap}
p_e^{ML}&=&\frac{1}{2}\int\delta_0(x)f_{y_s}(x|\xi=1)+\delta_1(x)f_{y_s}(x|\xi=0)\mathrm{d}x\\ \nonumber &\approx&\frac{1}{2}\int \min\{f_{\tilde{Z}}\Big(x-\frac{3\Lambda_s^{'}}{2}\tau^2|(\Lambda_s+\Lambda_0)(1-\tau)\Big),f_{\tilde{Z}}\Big(\tau-\frac{(\Lambda_s^{'}+\Lambda_0)}{2}\tau^2|\Lambda_0(1-\tau)\Big)\}\mathrm{d}x.
\ee
\subsection{Gaussian Approximation for Finite Sampling Rate in Linear Regime}\label{subsect.apgau}
We assume $k_s\dff\frac{\tau}{T_s}$ and $k_d\dff\frac{1}{\tau}$ are integers, i.e., $k_s$ samples in dead time $\tau$ and $k_sk_d$ samples in symbol duration $T_b$. For the output samples $\{Z_{1},\cdots,Z_{k_sk_d}\}$ in a symbol duration, the mean power $y_s^{k_s}$, defined as $y_s^{k_s}\dff\frac{1}{k_sk_d}\sum_{i=1}^{k_sk_d}Z_{i}$, is used for symbol detection. Noting that adjacent $k_s$ samples is related due to over-sampling rate with $h(t)=u(t)-u(t-\tau)$, we can rearrange the mean power as $y_s^{{k_s}}=y_{c}^{{k_s}}+y_{l}^{{k_s}}$, where $y_{c}^{{k_s}}$ and $y_{l}^{{k_s}}$ denote the current symbol component and previous symbol component with ${k_s}$ as the over-sampling rate, and corresponding the current symbol samples sequence and previous symbol samples sequence, defined as $\{Z_{1}^{c},\cdots,Z_{k_sk_d}^{c}\}$ and $\{Z_{1}^{l},\cdots,Z_{k_sk_d}^{l}\}$. Define random variable $G_{\Lambda}\dff\sum_{n=0}^{+\infty}\mathbbm{1}\{N_{\Lambda}=n\}G_{(n)}$, where random variable $N_{\Lambda}\sim Poi(\Lambda)$, $G_{(n)}\dff\sum_{i=1}^{n}G_i$ and $\{G_1,\cdots,G_n\}$ denote independent and identically distributed sequence of PMT gain. Since the photon arriving at the $[-\tau,0)$ would contribute to the mean power and $Z_i^l>0$ for $1\leq i\leq k_s-1$, we have
\be 
y_{l}^{{k_s}}&=&\frac{\sum_{i=1}^{k_sk_d}Z_i^{l}}{k_sk_d}=\frac{\sum_{i=1}^{{k_s}-1}i\sum_{n=0}^{+\infty}\mathbbm{1}\{N_{\frac{\tau\Lambda_{(1)}}{k_s}}=n\}G_{(n)}}{k_sk_d}=\frac{\sum_{i=1}^{{k_s}-1}iG_{\frac{\tau\Lambda_{(1)}}{k_s},i}}{k_sk_d},
\ee  
where $\Lambda_{(1)}$ denotes previous symbol photon arrival rate, and $\{G_{\frac{\tau\Lambda_{(1)}}{{k_s}},1},\cdots,G_{\frac{\tau\Lambda_{(1)}}{{k_s}},{k_s}-1}\}$ are i.i.d. with $G_{\frac{\tau\Lambda_{(1)}}{{k_s}}}$. 
Note that the mean power of the current symbol would leak into the next symbol with photons arriving at the $[1-\tau,1)$, then we have
\be 
y_{c}^{{k_s}}&=&\frac{\sum_{i=1}^{k_sk_d}Z_i^{c}}{k_sk_d}=\frac{1}{k_d}\sum_{n=0}^{+\infty}\mathbbm{1}\{N_{\frac{k_d-1}{k_d}\Lambda_{(0)}}=n\}G_{(n)}+\frac{1}{k_sk_d}\sum_{i=1}^{{k_s}-1}i\sum_{n=0}^{+\infty}\mathbbm{1}\{N_{\frac{\tau\Lambda_{(0)}}{k_s}}=n\}G_{(n)}\nonumber\\
&=&\frac{G_{\frac{k_d-1}{k_d}\Lambda_{(0)}}}{k_d}+\frac{\sum_{i=1}^{{k_s}-1}iG_{\frac{\tau\Lambda_{(0)}}{k_s},i}}{k_sk_d},
\ee 
where $\Lambda_{(0)}$ denotes the current symbol photon arrival rate , $G_{\frac{k_d-1}{k_d}\Lambda_{(0)}}$ and $G_{\frac{\tau\Lambda_{(0)}}{k_s},i}$ are mutually independent. 

As $Z_i$ and $Z_j$ are independent for $|i-j|\geq k_s$ and $k_d\gg1$, $y_s^{k_s}$ can be well approximated by Gaussian distribution with matched expectation and variance according to central limit theorem. Noting that $\mathbb{E}[G_\Lambda]=\Lambda$ and $\mathbb{D}[G_\Lambda]=\Lambda(1+2a^{-1})$, we have
\be 
\mathbb{E}[y^{k_s}_s|\Lambda_{(0)},\Lambda_{(1)}]&=&\frac{(k_d-1)\Lambda_{(0)}}{k_d^2}+\frac{(\Lambda_{(1)}+\Lambda_{(0)})\tau\sum_{i=1}^{k_s-1}i}{k_s^2k_d}\nonumber\\&=&\Lambda_{(0)}\tau(1-\tau)+(\Lambda_{(1)}+\Lambda_{(0)})\frac{\tau^2}{2}(1-k_s^{-1}),\\
\mathbb{D}[y^{k_s}_s|\Lambda_{(0)},\Lambda_{(1)}]&=&\frac{(k_d-1)\Lambda_{(0)}(1+2a^{-1})}{k_d^3}+\frac{(\Lambda_{(1)}+\Lambda_{(0)})\tau(1+2a^{-1})\sum_{i=1}^{k_s-1}i^2}{k_s^3k_d^2}\\ \nonumber &=&\Lambda_{(0)}\tau^2(1-\tau)(1+2a^{-1})+(\Lambda_{(1)}+\Lambda_{(0)})\frac{\tau^3}{6}(1+2a^{-1})(2-3k_s^{-1}+k_s^{-2}).
\ee 
It is seen that $\mathbb{E}[y_s^{k_s}|\Lambda_{(0)},\Lambda_{(1)}]$ and $\mathbb{D}[y_s^{k_s}|\Lambda_{(0)},\Lambda_{(1)}]$ increases and decreases with $k_s$, respectively. Assume $\Lambda_{(1)}$ is given during detecting $\Lambda_{(0)}$. To compare the error probability with different $k_s$ over-sampling rates, we normalize $y^{k_s}_s$ given $\xi=0$ such that $y^{k_s}_{s|\xi=0}$ can be approximated by normalized Gaussian distribution. For the scenario of $\xi = 1$, we have
\be 
\mu_{k_s}&\dff&\mathbb{E}_{\Lambda_{(1)}}\big[\mathbb{E}[y^{k_s}_{s}|\xi=1,\Lambda_{(1)}]\big]=\mathbb{E}_{\Lambda_{(1)}}\big[\frac{\mathbb{E}[y^{k_s}_s|\Lambda_{1},\Lambda_{(1)}]-\mathbb{E}[y^{k_s}_s|\Lambda_{0},\Lambda_{(1)}]}{\sqrt{\mathbb{D}[y^{k_s}_s|\Lambda_{0},\Lambda_{(1)}]}}\big] \nonumber\\&=&\frac{(\Lambda_{1}-\Lambda_{0})[1-\frac{\tau}{2}(1+k_s^{-1})]}{2\sqrt{1+2a^{-1}}}\{[\Lambda_{0}\big(1-\tau+\frac{\tau}{3}(2-3k_s^{-1}+k_s^{-2})\big)]^{-\frac{1}{2}}\nonumber\\&&+[\Lambda_{0}(1-\tau)+(\Lambda_{1}+\Lambda_{0})\frac{\tau}{6}(2-3k_s^{-1}+k_s^{-2})]^{-\frac{1}{2}}\},\\
\sigma^2_{k_s}&\dff&\mathbb{E}_{\Lambda_{(1)}}\big[\mathbb{D}[y^{k_s}_{s}|\xi=1,\Lambda_{(1)}]\big]=\mathbb{E}_{\Lambda_{(1)}}\big[\frac{\mathbb{D}[y^{k_s}_s|\Lambda_{1},\Lambda_{(1)}]}{\mathbb{D}[y^{k_s}_s|\Lambda_{0},\Lambda_{(1)}]}\big]\nonumber\\ \nonumber&=&1+\frac{(\Lambda_{1}-\Lambda_{0})\big(1-\tau+\frac{\tau}{6}(2-3k_s^{-1}+k_s^{-2})\big)}{2}\{[\Lambda_{0}\big(1-\tau+\frac{\tau}{3}(2-3k_s^{-1}+k_s^{-2})\big)]^{-1}\nonumber\\&&+[\Lambda_{0}(1-\tau)+(\Lambda_{0}+\Lambda_{1})\frac{\tau}{6}(2-\frac{3}{k_s}+\frac{1}{k_s^2})]^{-1}\},
\ee 
Then we have $f_{y^{{k_s}}_{s}|\xi=1}\sim\mathcal{N}(\mu_{k_s},\sigma^2_{k_s})$, where $\mathcal{N}(\mu_{k_s},\sigma^2_{k_s})$ denotes the Gaussian distribution with expectation $\mu_{k_s}$, variance $\sigma^2_{k_s}$. Based on ML detection, we have the following approximate threshold $\gamma_{1a}^{{k_s}}$,
\be 
\gamma_{1a}^{{k_s}}=\frac{-\mu_{{k_s}}+\sqrt{\mu_{{k_s}}^2+(\sigma^2_{{k_s}}-1)(\mu_{{k_s}}^2+\sigma^2_{{k_s}}\ln\sigma^2_{{k_s}})}}{\sigma^2_{{k_s}}-1};
\ee 
and thus the detection error probability $p_{e,{k_s}}^{wa}$ is given by
\be 
p_{e,{k_s}}^{wa}=\frac{1}{2}\Big(Q(\gamma_{1a}^{{k_s}})+Q(\frac{\mu_{{k_s}}-\gamma_{1a}^{{k_s}}}{\sigma_{{k_s}}})\Big).
\ee 
%---------------------------------------------------------------------------------------
\subsection{Infinite Sampling Rate in the Non-linear Region}
The numbers of arrived photons $N$ and the corresponding photon arrival time $\mathbf{T}^N=(t_1,\cdots,t_N)$ is a sufficient statistic of Poisson channel. Via eliminating the order of arrival time with i.i.d sequence $\{t_1,\cdots,t_n\}$ provided $N=n$, we have the following conditional sample function density:
\be 
f_{(N,\mathbf{\hat{T}}^N)|\lambda}(n,t^n)=\left\{\begin{array}{ll}
	e^{-\lambda},&n=0;\\
	e^{-\lambda}\frac{\lambda^n}{n!},&t_i\in[0,1],i=1,\cdots,n.
\end{array}\right.
\ee 
The output signal is given by $y(t)=C\big(\sum\limits_{t_i\in[t-\tau,t)}G_i\big)$, where $G_i$ denotes the gain of photons arrived at time $t_i$. The average power $y_s$ is given by
\be 
y_s=\int_{0}^{1}y(t)\mathrm{d}t=\sum_{n=0}^{+\infty}e^{-\lambda}\frac{\lambda^n}{n!}\int_{0}^{1}\int_{[0,1]^n} C\big(\sum\limits_{t_i\in[t-\tau,t)}G_i\big)\mathrm{d}t_1^n\mathrm{d}t.
\ee 
Then, the detection error probability based on ML detection is given by,
\be
p_e^{NML}&=&\int\min\{f_{y_s}(x|\xi=1),f_{y_s}(x|\xi=0)\}\mathrm{d}x.
\ee

\section{Mean power detection and photon counting detection for under-sampling }\label{sec.detefi}
For practical interest, we consider the scenario of lower sampling rate, where the sampling interval $T_s\geq\tau$, and further assume $L=\frac{1}{T_s}$ is an integer. Thus, the sample values $\tilde{Z}_1^{L}=\{\tilde{Z}_1,\cdots,\tilde{Z}_{L}\}$ are independent and identically distributed. We analyze the probability density function of samples considering the following two cases depending on the signal intensity.

\textbf{Linear Case:} Assume $C(x)=x$ if signal intensity does not reach the threshold for the nonlinear regime. According to Equation (\ref{eq.MGFnorg}), the MGF and probability density of $\tilde{Z}^{l}$ are given by,
\be 
M_{\tilde{Z}^{l}|\lambda}(\omega)&=&e^{\lambda\big(\exp(-\frac{\omega}{1+a^{-1}\omega})-1\big)},\\
f_{\tilde{Z}^{l}|\lambda}(z)&=&e^{\lambda(e^{-a}-1)}\delta(z)+a\sqrt{\frac{1}{z}}e^{-(\lambda+az)}\sum_{n=0}^{\infty}\frac{\sqrt{n}(\lambda e^{-a})^n}{n!}I_1(2a\sqrt{nz}).
\ee 

\textbf{Non-linear Case:} If the signal intensity increases beyond a threshold, the effect of anode non-linearity emerges. Assume nonlinear function $C(\cdot)$ satisfies conditions in Section \ref{sec.classPMT} and increases in $[0,x_s]$ firstly and then decreases in $[x_s,\infty)$ corresponding to current saturation and supersaturation for a certain threshold $x_s$. Note that $C(x)=C_u(x)\mathbbm{1}\{x\leq x_s\}+C_d(x)\mathbbm{1}\{x>x_s\}$, where $C_u(\cdot)$ and $C_d(\cdot)$ denote the strictly increasing and decreasing parts of $C(\cdot)$, respectively. Defining $\tilde Z^{nl} \dff C(\tilde Z^{l})$, we have the following probability density of $\tilde{Z}^{nl}$,
\be\label{eq.nonlinaer}
f_{\tilde{Z}^{nl}|\lambda}(z)=\frac{f_{\tilde{Z}^{l}|\lambda}(C_u^{-1}(z))}{C_u^{'}(C_u^{-1}(z))}-\frac{f_{\tilde{Z}^{l}|\lambda}(C_d^{-1}(z))}{C_d^{'}(C_d^{-1}(z))},
\ee 
where $C_u^{-1}(\cdot)$ and $C_d^{-1}(\cdot)$ are inverse functions of $C_u(\cdot)$ and $C_d(\cdot)$, respectively. Equation (\ref{eq.nonlinaer}) can be employed to obtain $f_{\tilde{Z}^{nl}|\lambda}(z)$ in simulation.

In the remainder of this Section, we adopt $f_{\tilde{Z}|\lambda}(x)$ to denote $f_{\tilde{Z}^{l}|\lambda}(z)$ in the linear regime and $f_{\tilde{Z}^{nl}|\lambda}(z)$ in the non-linear regime. The joint probability density of samples in a symbol duration is given as follows
\be 
f_{Z_1,\cdots,Z_{L}|\lambda}(z_1,\cdots,z_{L})=\prod_{k=1}^{L}f_{\tilde Z|\lambda}(z_k).
\ee
\subsection{Photon Counting Detection}
For mutually independent samples, it's efficient to count by threshold rather than rising edge. Let $\gamma_2$ and $\hat{N}$ denote decision threshold and corresponding number of detected photons. Defining $F_k\dff\mathbbm{1}\{z_k\geq\gamma_2\}$, we have $\hat{N}=\sum_{k=1}^{L}F_k$. Obviously, the distribution of $\hat{N}$ follows binomial distribution $\mathbb{B}(L,p)$, where $p$ is determined by the probability of exceeding threshold $\gamma_2$
\be 
p_0\triangleq\mathbb{P}(\tilde{z}_k\geq\gamma_2|\lambda_0)=\int_{\gamma_2}^{\infty}f_{\tilde{Z}|\lambda_0}(z)\mathrm{d}z,\\
p_1\triangleq\mathbb{P}(\tilde{z}_k\geq\gamma_2|\lambda_1)=\int_{\gamma_2}^{\infty}f_{\tilde{Z}|\lambda_1}(z)\mathrm{d}z,
\ee
where $\lambda_0=\Lambda_0\tau$ and $\lambda_1=\Lambda_1\tau$. Defining $P_0^{B}=\mathbb{B}(L,p_0)$ and $P_1^{B}=\mathbb{B}(L,p_1)$ as the binomial distributions of $\hat{N}$ for symbol $0$ and symbol $1$, respectively, we have the following KL distance:
\be 
KL(P_0^{B}||P_1^{B})=L(p_0\log\frac{p_0}{p_1}+(1-p_0)\log\frac{1-p_0}{1-p_1}),\\
KL(P_1^{B}||P_0^{B})=L(p_1\log\frac{p_1}{p_0}+(1-p_1)\log\frac{1-p_1}{1-p_0}).
\ee 

According to Chernoff-Stein Lemma \cite{cover2012elements}, we pursue the optimal decision threshold $\gamma_2^{*}$ that maximizes the minimum of the above two KL distances, defined as follows,
\be\label{thres2opta}
\gamma_2^{*}=\arg \max\limits_{\gamma_2}\min\{KL(P_0^{B}||P_1^{B}),KL(P_1^{B}||P_0^{B})\}.
\ee 
Noting that $p_0<p_1$ is needed to maintain reliable communication, we have
\be 
KL(P_0^{B}||P_1^{B})-KL(P_1^{B}||P_0 ^{B})=(p_0-p_1)\log\frac{p_0(1-p_0)}{p_1(1-p_1)}\gtreqqless0,\text{ if }p_0+p_1\lesseqqgtr1.
\ee 
Based on above statement, Problem (\ref{thres2opta}) is equivalent to the following optimization problem:
\begin{equation}
\begin{aligned}
& \underset{\gamma_2}{\text{max}}
& &\min\{KL(P_0^{B}||P_1^{B}),KL(P_1^{B}||P_0^{B})\},\\
& \text{s.t.}
& &p_0=\int_{\gamma_2}^{\infty}f_{\tilde{Z}|\lambda_0}(z)\mathrm{d}z;\\
&&&p_1=\int_{\gamma_2}^{\infty}f_{\tilde{Z}|\lambda_1}(z)\mathrm{d}z.
\end{aligned}
\end{equation}

The above optimization problem can be solved via exhaustive search over $\gamma_2$.
\iffalse
 Moreover, we propose two heuristic suboptimal decision threshold as follows:
\be 
\gamma_{2a}^{*}=\arg \max\limits_{\gamma_2}p_1-p_0, \\
p_1(\gamma_{2b}^{*})+p_0(\gamma_{2b}^{*})=1.
\ee
\fi
Based on the optimized decision threshold $\gamma_2^{*}$, two likelihood functions are proposed as $P_0^B\dff\mathbb{B}(L,p_0(\gamma_2^{*}))$ and $P_1^B\dff\mathbb{B}(L,p_1(\gamma_2^{*}))$. Defining $\hat{n}$ as the number of detected photons, the log-likelihood ratio (LLR) is given as follows,
\be LLR_C\overset{\bigtriangleup}{=}\hat{n}\ln\frac{p_1}{p_0}+(L-\hat{n})\ln\frac{1-p_1}{1-p_0}.
\ee
We can obtain the following counting threshold, denoted as $\hat{n}_{th}$,
\be 
\hat{n}_{th}=\lfloor Lp_{th} \rfloor,
\ee
where $p_{th}=\frac{\log\frac{1-p_0}{1-p_1}}{\log\frac{p_1}{p_0}+\log\frac{1-p_0}{1-p_1}}$, such that symbol 1 is detected if $\hat{n}>\hat{n}_{th}$ and symbol 0 is detected otherwise. The detection error probability is given by
\be\label{eq.BERpcd2}
p_e^{c}=\frac{1}{2}\big(\sum_{n=0}^{\hat{n}_{th}}P_1^{B}(n)+\sum_{n=\hat{n}_{th}+1}^{L}P_0^{B}(n)\big).
\ee 
We provide the following bounds on threshold $p_{th}$.
\begin{lemma}\label{lemma.inq}
	$\frac{\hat{n}_{th}}{L}\leq p_{th}<p_1$ and $p_0<p_{th}\leq\frac{\hat{n}_{th}+1}{L}$.
	\begin{proof}
		Define function $s(x)=x\ln\frac{x}{x_0}-(1-x)\ln\frac{1-x_0}{1-x}$, for $x\geq x_0$ and $x_0<1$. Then, we have $s^{'}(x)=\ln\frac{x}{1-x}-\ln\frac{x_0}{1-x_0}\geq s^{'}(x_0)=0$, and thus $s(x)\geq0$ for $x\geq x_0$. Letting $x_0=p_0,x=p_1$ and $x_0=1-p_1,x=1-p_0$, we have $p_{th}<p_1$ and $p_0<p_{th}$.  
	\end{proof}
\end{lemma}

According to Chernoff theorem \cite{cover2012elements}, the optimum achievable exponent in terms of $p_e^c$ is the Chernoff information $C(P_1,P_0)=KL(P_{\lambda^{*}}||P_1)=KL(P_{\lambda^{*}}||P_0)$ with $P_{\lambda}=\frac{P_1^{\lambda}(x)P_0^{1-\lambda}(x)}{\sum\limits_{a\in\{0,1\}}P_1^{\lambda}(a)P_0^{1-\lambda}(a)}$, and $\lambda^{*}$ determined by $KL(P_{\lambda^{*}}||P_1)=KL(P_{\lambda^{*}}||P_0)$. Define $P_{th}$ as binomial distribution $\mathbb{B}(1,p_{th})$ for abbreviation, it can be verified readily that $KL(P_{th}||P_1)=KL(P_{th}||P_0)$ and $P_{th}\overset{d}{=}P_{\lambda^{*}}$ with $\lambda^{*}=\frac{\ln\big(\frac{1-p_0}{p_0}\ln\frac{1-p_0}{1-p_1}/\ln\frac{p_1}{p_0}\big)}{\ln\frac{p_1}{1-p_1}-\ln\frac{p_0}{1-p_0}}$.

\subsection{Mean Power Detection}
Defining $\tilde{X}_s=\frac{1}{L}\sum_{i=1}^{L}\tilde{Z}_i$ as the average power and $\gamma_1$ as the decision threshold, we have the following MGF of $\tilde{X}_s$ and probability density function,
\be 
M_{\tilde{X}_s|\lambda}(\omega)&=&\prod_{i=1}^{L}e^{\lambda\big(\exp(-\frac{\omega/L}{1+a^{-1}\omega/L})-1\big)}=e^{L\lambda\big(\exp(-\frac{\omega}{L+a^{-1}\omega})-1\big)},\\
f_{\tilde{X}_s|\lambda}(y)&=&Lf_{\tilde{Z}|L\lambda}(Ly),
\ee 
and thus symbol $\xi=1$ is detected if $LLR_W=\log\frac{f_{\tilde{Z}|L\lambda_1}(Ly)}{f_{\tilde{Z}|L\lambda_0}(Ly)}>0$, and symbol $\xi=0$ is detected otherwise. The detection error probability is given by
\be\label{eq.BERavd}
p_e^{L}=\frac{1}{2}\int_{0}^{\infty}\min\{f_{\tilde{Z}|L\lambda_0}(y),f_{\tilde{Z}|L\lambda_1}(y)\}\mathrm{d}y.
\ee 
Then it is seen that the performance of mean power detection given the number of samples $L$ is equivalent to that with single-rate sampling with $L$ times of signal and background photon arrival rates.

For reliable communication, samples sequence length $L$ requires sufficiently large and thus $\tilde{X}_s$ can be approximated by Gaussian distribution according to central limit theorem. Strict proof of asymptotically Gaussian of $\tilde{X}_s$ for large $L$ is similar to Lemma \ref{lem.gauap} and omitted here. In the following, Gaussian approximation is adopted to derive approximate closed-form detection error probability $p_e^{La}$. 

Note that the expectation and variance of $\tilde{X}_s$ are given by $\mathbb{E}[\tilde{X}_s|\xi=i]=\lambda_i$ and $\mathbb{D}[\tilde{X}_s|\xi=i]=\frac{\lambda_i(1+2a^{-1})}{L}$ for $i=0,1$. Similarly to Section \ref{subsect.apgau}, we normalize $\tilde{X}_s$ given $\xi=0$ such that it can be approximated by normalized Gaussian distribution. Then, we have the following approximate threshold $\gamma_{MPD}^{{La}}$,
\be 
\gamma_{MPD}^{{La}}=\frac{-\mu_{La}+\sqrt{\mu_{La}^2+(\sigma^2_{La}-1)(\mu_{La}^2+\sigma^2_{La}\ln\sigma^2_{La})}}{\sigma^2_{La}-1};
\ee 
where $\mu_{La}=\frac{\lambda-1-\lambda_0}{\sqrt{\lambda_0(1+2a^{-1})/L}}$ and $\sigma^2_{La}=\frac{\lambda_1}{\lambda_0}$. Thus, the detection error probability $p_{e,{k_s}}^{wa}$ is given by
\be \label{eq.BERgauapp}
p_e^{La}=\frac{1}{2}\Big(Q(\gamma_{MPD}^{{La}})+Q(\frac{\mu_{{La}}-\gamma_{MPD}^{{La}}}{\sigma_{{La}}})\Big).
\ee

\iffalse
\subsection{Joint detection}
A natural idea is to combine these two types of detection. The combined symbol detection is based on the weighted summation of LLRs of the average power detection and the photon counting detection as follows
\be 
LLR_{CW}=\alpha LLR_{C}+(1-\alpha) LLR_{W}
\ee 
where $\alpha$ is the weight of photon counting detection and could be optimized to minimal the error probability of combination detection. Define decision region as follow:
\be 
\delta_0(\mathbf{\tilde{Z}})=\mathbbm{1}\{LLR_{CW}(\mathbf{\tilde{Z}},\alpha)<0\},\\
\delta_1(\mathbf{\tilde{Z}})=\mathbbm{1}\{LLR_{CW}(\mathbf{\tilde{Z}},\alpha)\geq0\}.
\ee
and then we have error probability given by
\be 
p_e^{cw}=\frac{1}{2}\big(\int\delta_0(\mathbf{\tilde{Z}})P_{\mathbf{\tilde{Z}}}(\mathbf{\tilde{Z}}|\lambda_1)\mathrm{d}\mathbf{\tilde{Z}}+\int\delta_1(\mathbf{\tilde{Z}})P_{\mathbf{\tilde{Z}}}(\mathbf{\tilde{Z}}|\lambda_0)\mathrm{d}\mathbf{\tilde{Z}}\big)
\ee 
\fi
\section{Experimental and numerical results}\label{sec.NumericalResults}
In experiment, a small threshold is set to filter negligible thermal noise and detect the pulses under weak illuminate aiming to avoid pulse merge. The PMT gain samples refer to the amplitude of the pulses. Standard Gaussian kernel density estimation is adopted to obtain the estimated probability density with 1540 samples. Figure \ref{fig.pdfexp} shows the normalized gain probability distribution with mean one obtained from the asymmetric model, experiment and Gaussian approximation. The gain probability distribution is non-negative and asymmetric, where the asymmetric distribution model is more accurate than the Gaussian approximation.

In simulation, the PMT output signal is generated via the pulses for the photon arrival process and a non-linear function. Consider a PMT receiver with asymmetric shot noise and negligible thermal noise. We adopt the following system parameters: symbol rate $1$Msps; mean PMT gain $3\times10^7$; dead time $20ns$; background photon rate $100000s^{-1}$, such that the normalized dead time is $0.02$ and the normalized background photon rate is $0.1$.

Figures \ref{fig.opmulam}(a) and \ref{fig.opmulam}(b) plot the optimal duty cycle against peak-power $\Lambda_s$, with single and double sampling rates for different background photon intensities to maximize $I(b;\tilde{Z})$ according to Equations (\ref{eq.suboptmu}) and (\ref{eq.solumulti}), respectively, from both simulations and numerical computations. It is seen that the optimal duty cycles from theoretical derivations and simulations are very close to each other. Moreover, the proposed suboptimal duty cycle corresponding to $\Lambda_0=0$ is very close to the optimum one for high $\Lambda_s$ or low $\Lambda_0$. 

Figure \ref{fig.threesignal2} shows the typical signals in the three signal regimes from simulations, along with the trend of pulse merge and saturation as $\Lambda$ increases. As for threshold-based criterion, setting $l_{max}=2.4$ and $\epsilon=0.015$ for the threshold between waveform regime and transition regime, we have $\Lambda^{(1)}_{th}=50$ and $\Lambda^{(2)}_{th}=518.425$ with the corresponding simulated signal waveforms shown in Figures \ref{fig.thresholdsignal1} and \ref{fig.thresholdsignal2}, respectively. It is seen that the proposed model can well characterize the signals and the regime transitions.

Figure \ref{fig.pdflam} shows the probability density function (pdf) of the normalized PMT signal and Gaussian fitting results via matching the first and second moments, which shows the accuracy of Gaussian approximation. Moreover, Gaussian distribution shows better approximation performance for large peak power, which is consistent with asymptotic Gaussian approximation from Lemma \ref{lem.gauap}.

Figures \ref{fig.meanvar2} and \ref{fig.meanvar1} show the fitted non-linear function, along with the normalized mean and standard deviation (STD) of the received signals from experiments and simulations. The determination of fitted non-linear function is elaborated in Appendix \ref{appd.deter}. In Broyden-Fletcher-Goldfarb-Shanno (BFGS) algorithm, we set the length of experimental data $M_1=31$, the length of variable $M_2=201$ with value from $0$ to $10$ with interval $0.05$, the maximum number of iterations $100$ and initial non-linear function as linear function. Armijo criterion is adopted to determine the step size in the search. Such non-linear function is consistent with the trend of the mean of PMT output signals, and the mean and STD from simulations show the same trend as those from experiments. 

%Based on Gaussian approximation, the thresholds $\lambda^{(2)}_{th}$ for different $l_{max}$s are shown in Figure \ref{fig.threpro}. It's seen that the threshold $\lambda^{(2)}_{th}$ decreses with probability $\epsilon$ and increses with $\l_{max}$. 
Figure \ref{fig.MPDover} shows the BER of MPD with different over-sampling rates from simulations as well as from Gaussian approximation and Equation (\ref{eq.BERintap}). It is seen that the improvement of MPD brought by higher samping rate is not significant, and Gaussian approximation shows larger gap for over-sampling and higher $\lambda_s$ due to inaccurate tail distribution estimation. The gap between Equation (\ref{eq.BERintap}) and simulation can be explained by the omitted but non-negligible high order terms.

Figure \ref{fig.BERlamdown} plots the BER of MPD and PCD against signal arrival rate $\Lambda_s$ for $L=50, 25, 10$ from both Monte-Carlo simulations, theoretical analysis and Gaussian approximation analysis according to Equations (\ref{eq.BERpcd2}), (\ref{eq.BERavd}) and (\ref{eq.BERgauapp}), respectively, in the linear regime. In Monte-Carlo simulations, we set the number of random symbols to be $5\times10^4$ and adopt ML detection. It is seen that the BER of threshold-based PCD is lower than or approximately the same as that of MPD. 

Figure \ref{fig.BERlamdown2} plots the BER of MPD and PCD against signal arrival rate $\Lambda_s$ for $L=50, 25, 10$ and for non-linear function $C(x)$ shown in Figure \ref{fig.meanvar2} with linear interpolation, and Gaussian thermal noise power $\sigma_{0}^2=10^{-3}$ from simulations with $5\times10^5$ random data symbols based on the optimal threshold. It is seen that the BER of PCD is higher than that of MPD in the low SNR regime and lower in the high SNR regime. It is justified by the fact that noise can be removed more effectively via the average operation compared with the hard-decision operation for low SNR, and the average operation results in larger performance loss compared with the hard-decision operation for high SNR.
\begin{figure}
		\centering
		{\includegraphics[angle=0, width=0.8\textwidth]{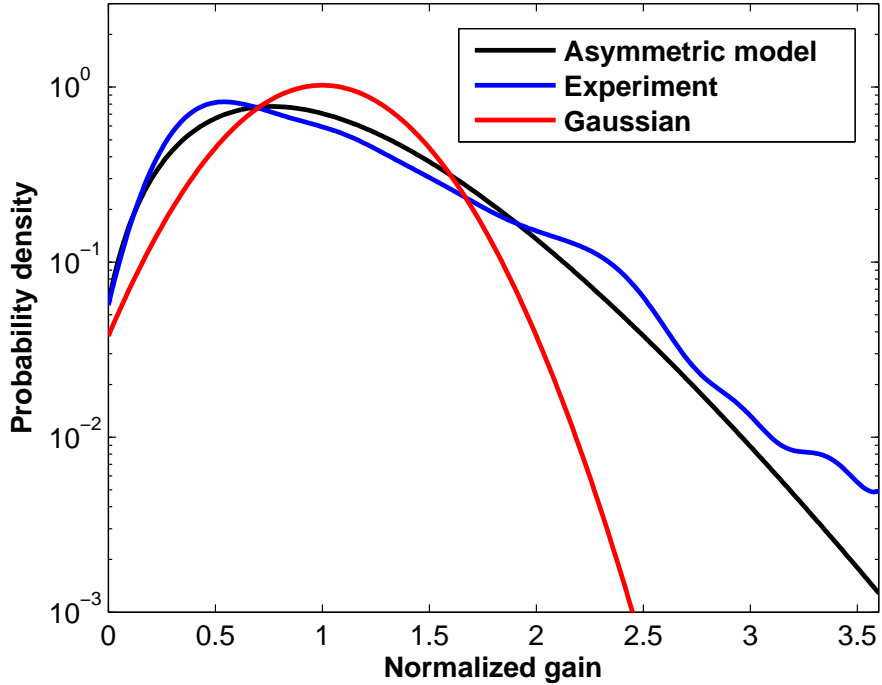}}
		\caption{The probability distribution of asymmetric model, estimation in experiment and Gaussian approximation.}
		\label{fig.pdfexp}
\end{figure}
\begin{figure}[htbp]
	\centering
	\subfigure[]{\includegraphics[width =3in]{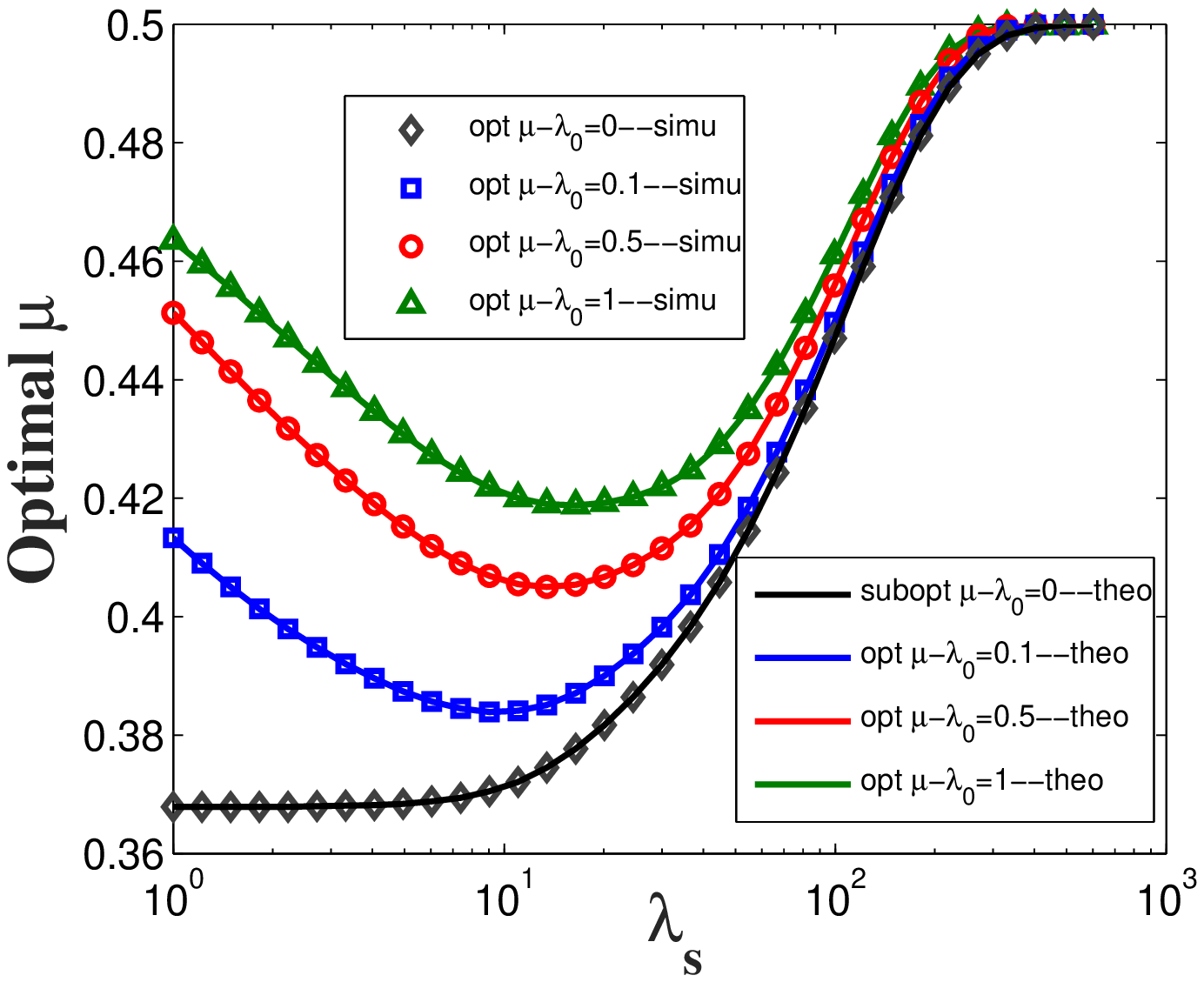}}
	\subfigure[]{\includegraphics[width =3in]{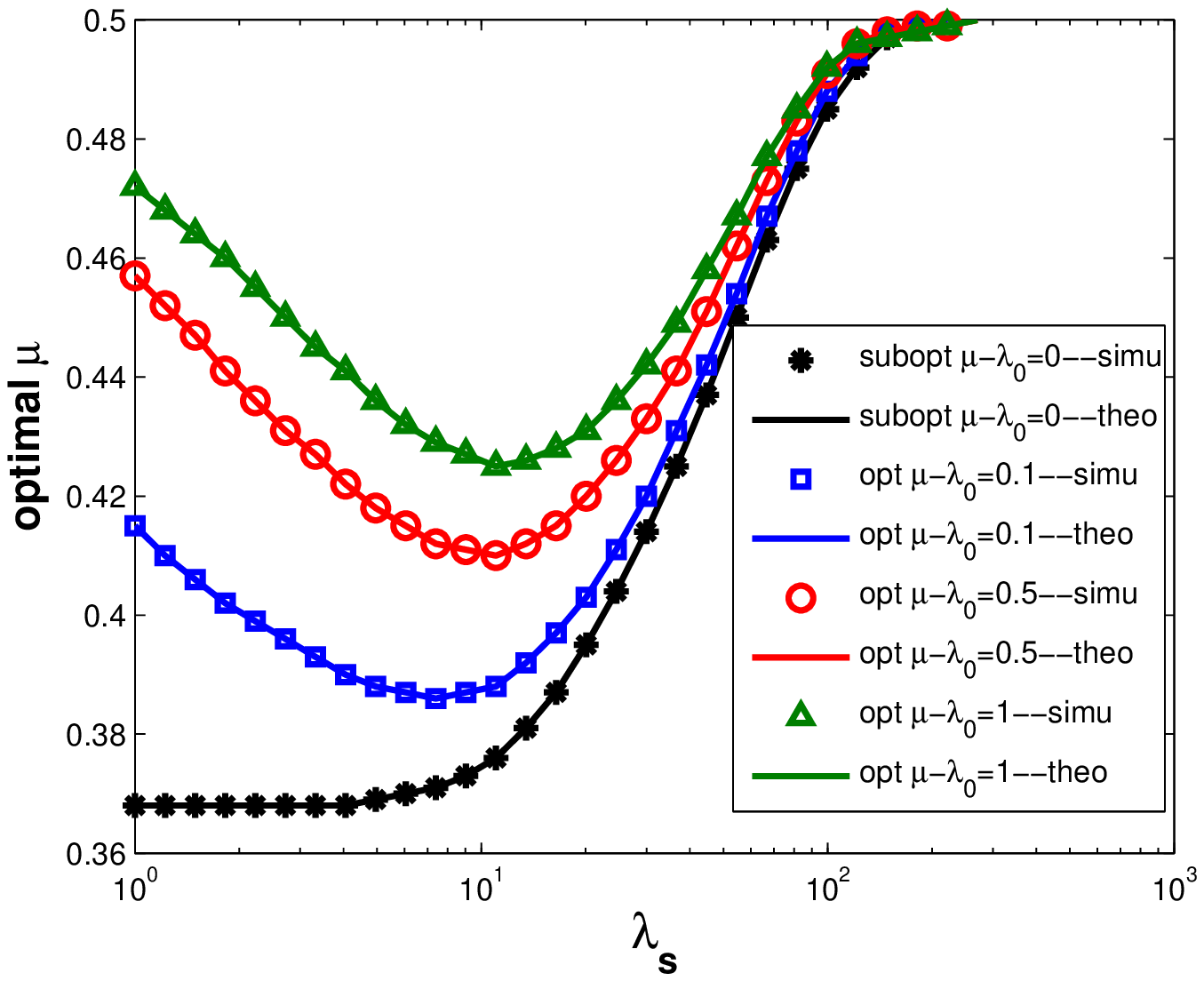}}
	\caption{The optimal duty cycle versus $\lambda_s$ from simulation and theoretical analysis for different $\lambda_0$ with single-rate sampling $T_s=T_b$ in (a) and double-rate sampling $T_s=\frac{T_b}{2}$ in (b).}
	\label{fig.opmulam}
\end{figure}
\begin{figure}	
	\begin{minipage}[t]{0.45\textwidth}
		\centering
		{\includegraphics[angle=0, width=1.0\textwidth]{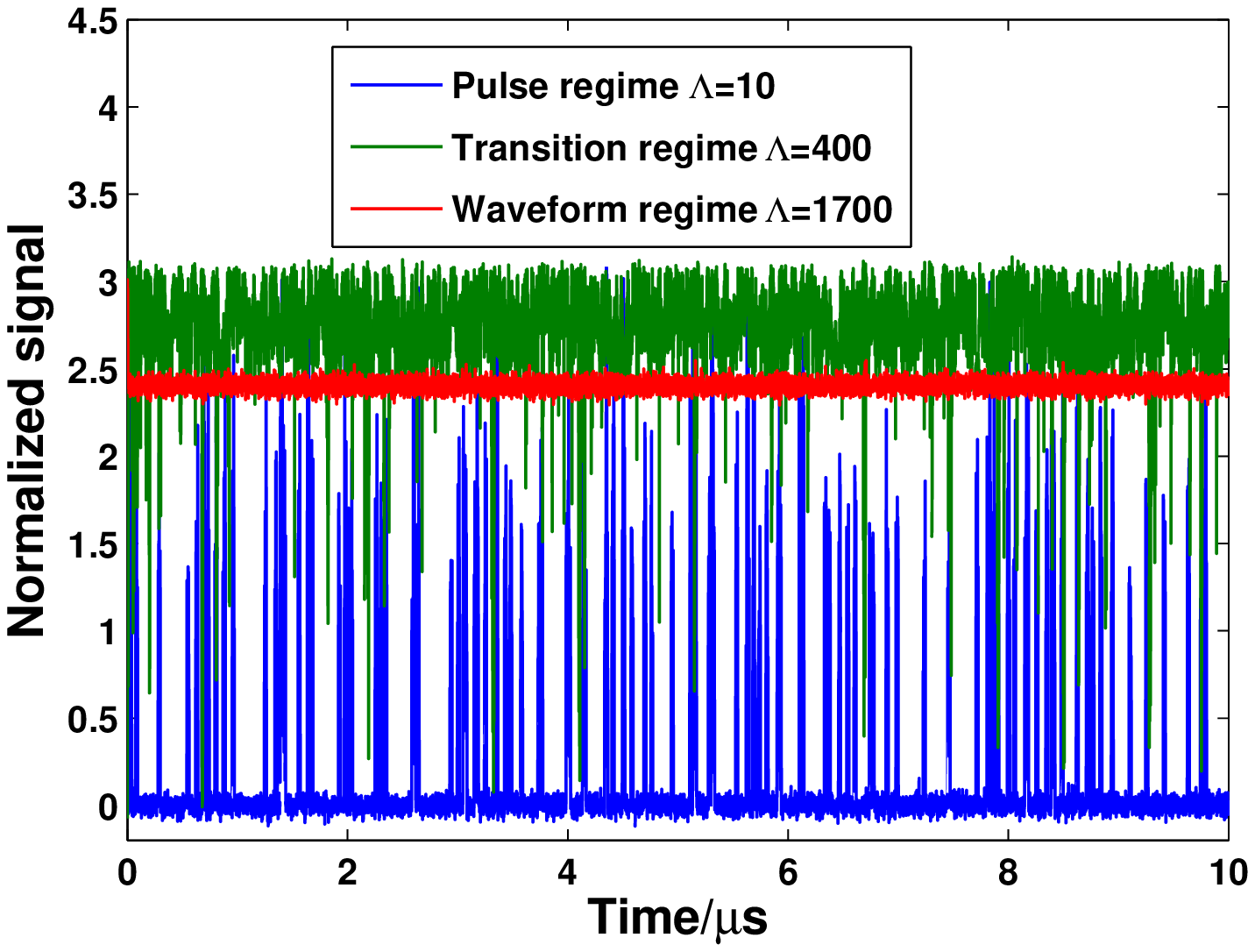}}
		\caption{The typical signal waveforms of three PMT working regimes from simulations.}
		\label{fig.threesignal2}
	\end{minipage}
	\begin{minipage}[t]{0.45\textwidth}
		\centering
		{\includegraphics[angle=0, width=1.0\textwidth]{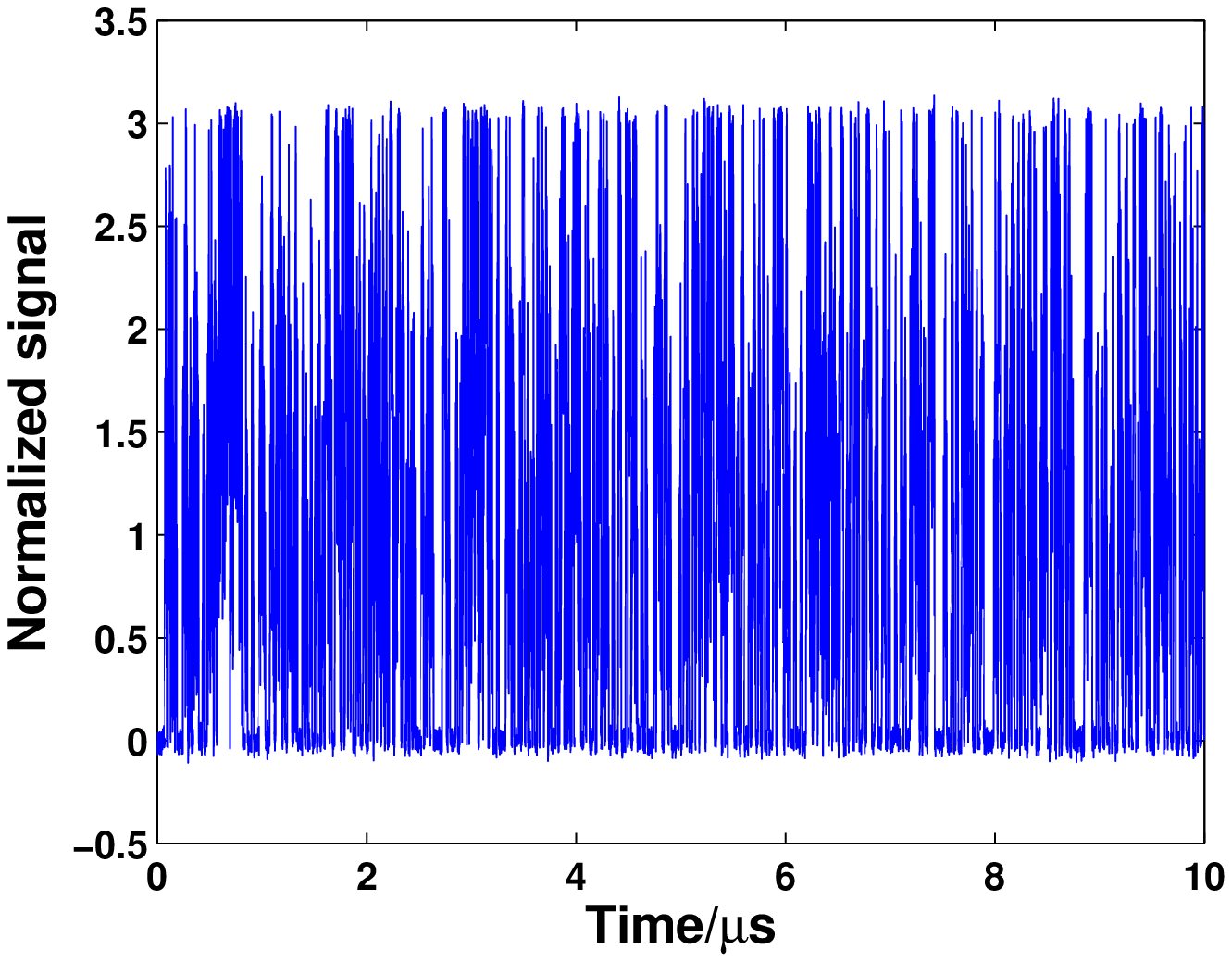}}
		\caption{The waveforms for the signal intensity at the threshold between pulse regime and transition regime.}
		\label{fig.thresholdsignal1}
	\end{minipage}
\end{figure}
\iffalse
\begin{figure}[htb]
	\setlength{\abovecaptionskip}{-0.3cm} %缩小caption和图像之间的距离
	\setlength{\belowcaptionskip}{0cm}
	\centering
	{\includegraphics[width =4.5in]{PDFexp2.eps}}
	\caption{The probability distribution of asymmetric model, estimation in experiment and Gauss.}
	\label{fig.pdfexp}
\end{figure}
\begin{figure}[htbp]
	\centering
	{\includegraphics[width =4.5in]{threesig_simu.eps}}
	\caption{The typical signal waveforms of three PMT working regimes from simulations.}
	\label{fig.threesignal2}
\end{figure}
\fi
\begin{figure}		
	\begin{minipage}[t]{0.45\textwidth}
		\centering
		{\includegraphics[angle=0, width=1.0\textwidth]{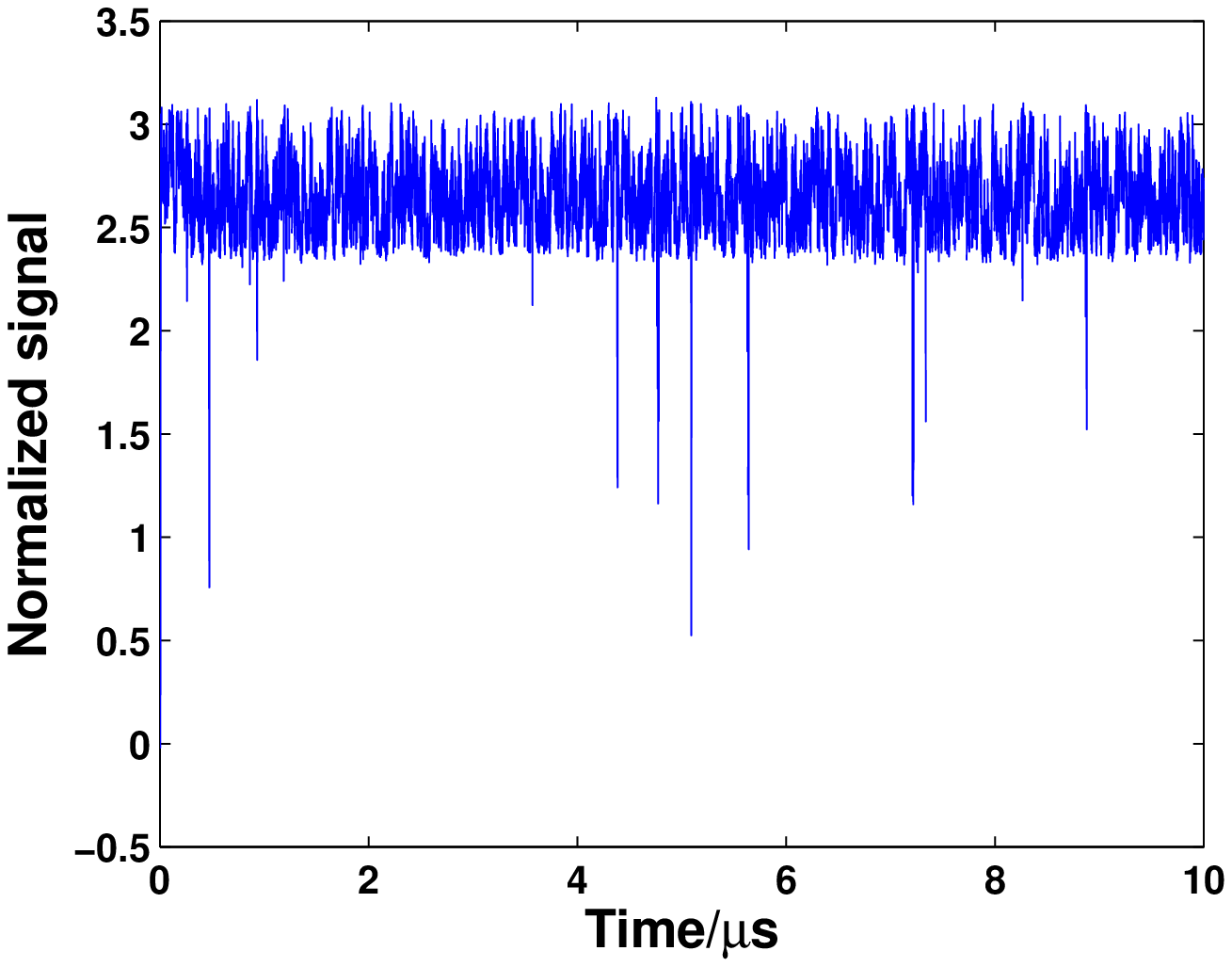}}
		\caption{The waveforms for the signal intensity at the threshold between transition regime and waveform regime.}
		\label{fig.thresholdsignal2}
	\end{minipage}
	\begin{minipage}[t]{0.45\textwidth}
		\centering
		{\includegraphics[angle=0, width=1.0\textwidth]{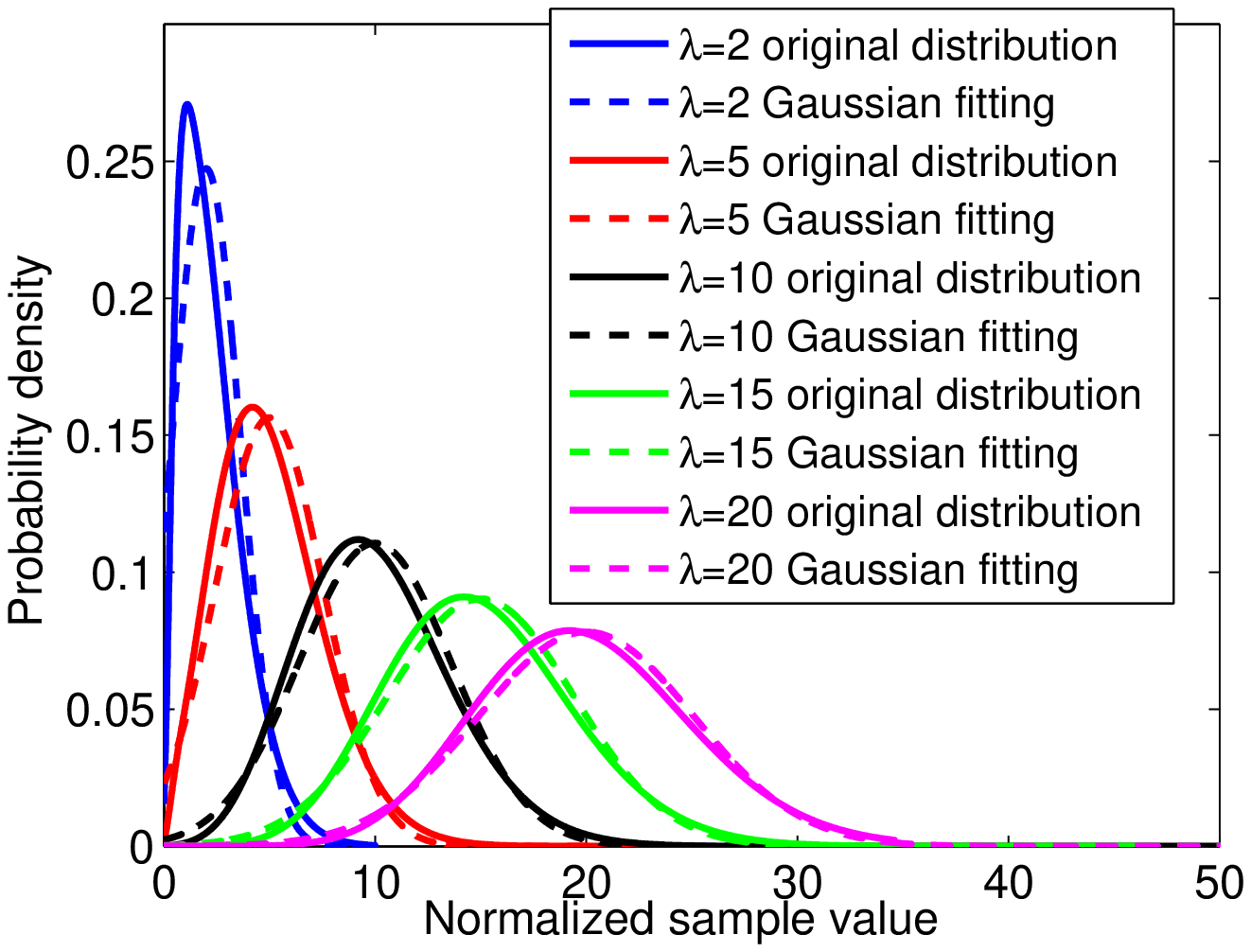}}
		\caption{The PMT gain PDFs and the Gaussian approximations for different $\lambda$ values.}
		\label{fig.pdflam}
	\end{minipage}
\end{figure}

\begin{figure}	
	\begin{minipage}[t]{0.45\textwidth}
		\centering
		{\includegraphics[angle=0, width=1.0\textwidth]{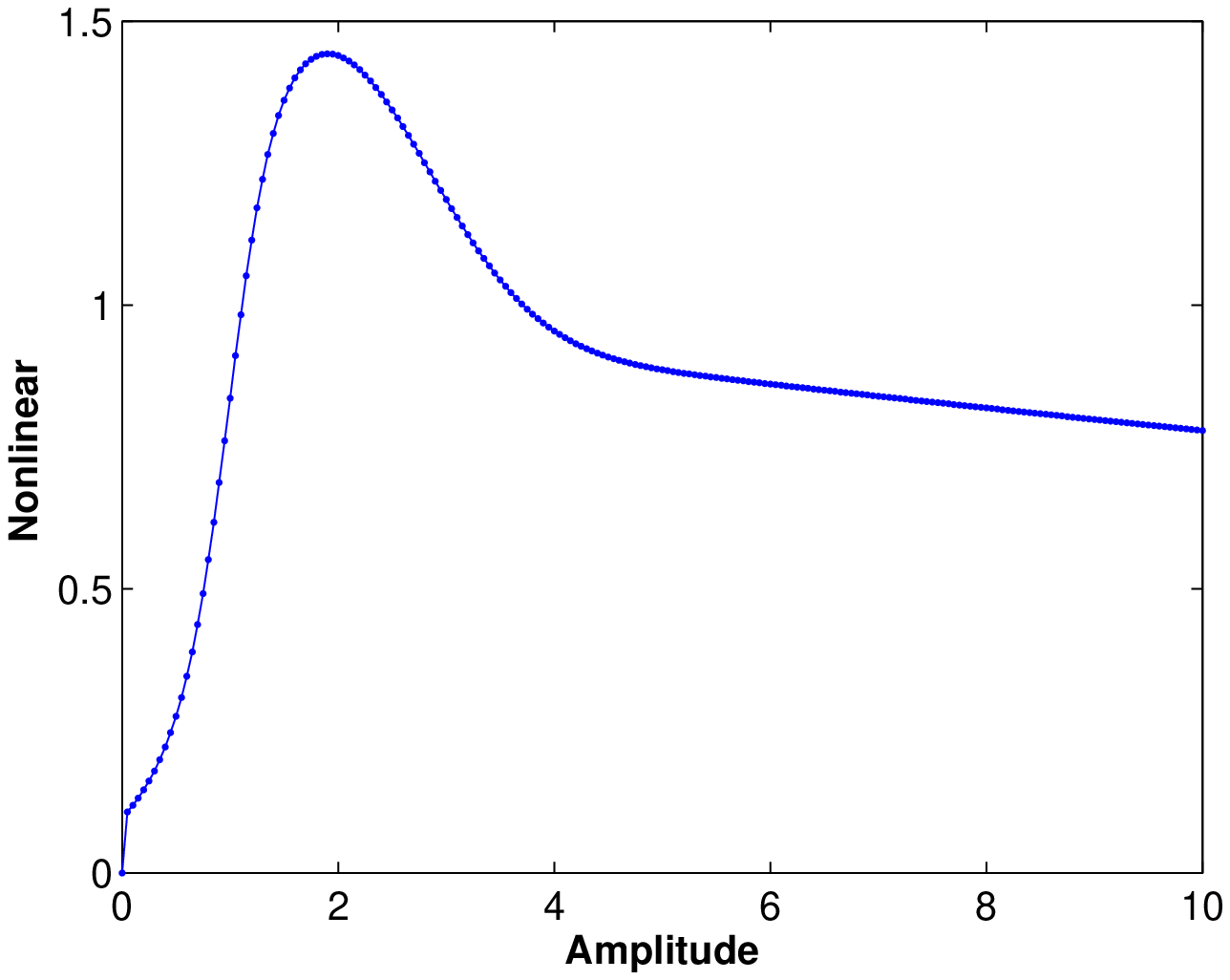}}
		\caption{The optimal non-linear function by BFGS algorithm.}
		\label{fig.meanvar2}
	\end{minipage}		
	\begin{minipage}[t]{0.45\textwidth}
		\centering
		{\includegraphics[angle=0, width=1.0\textwidth]{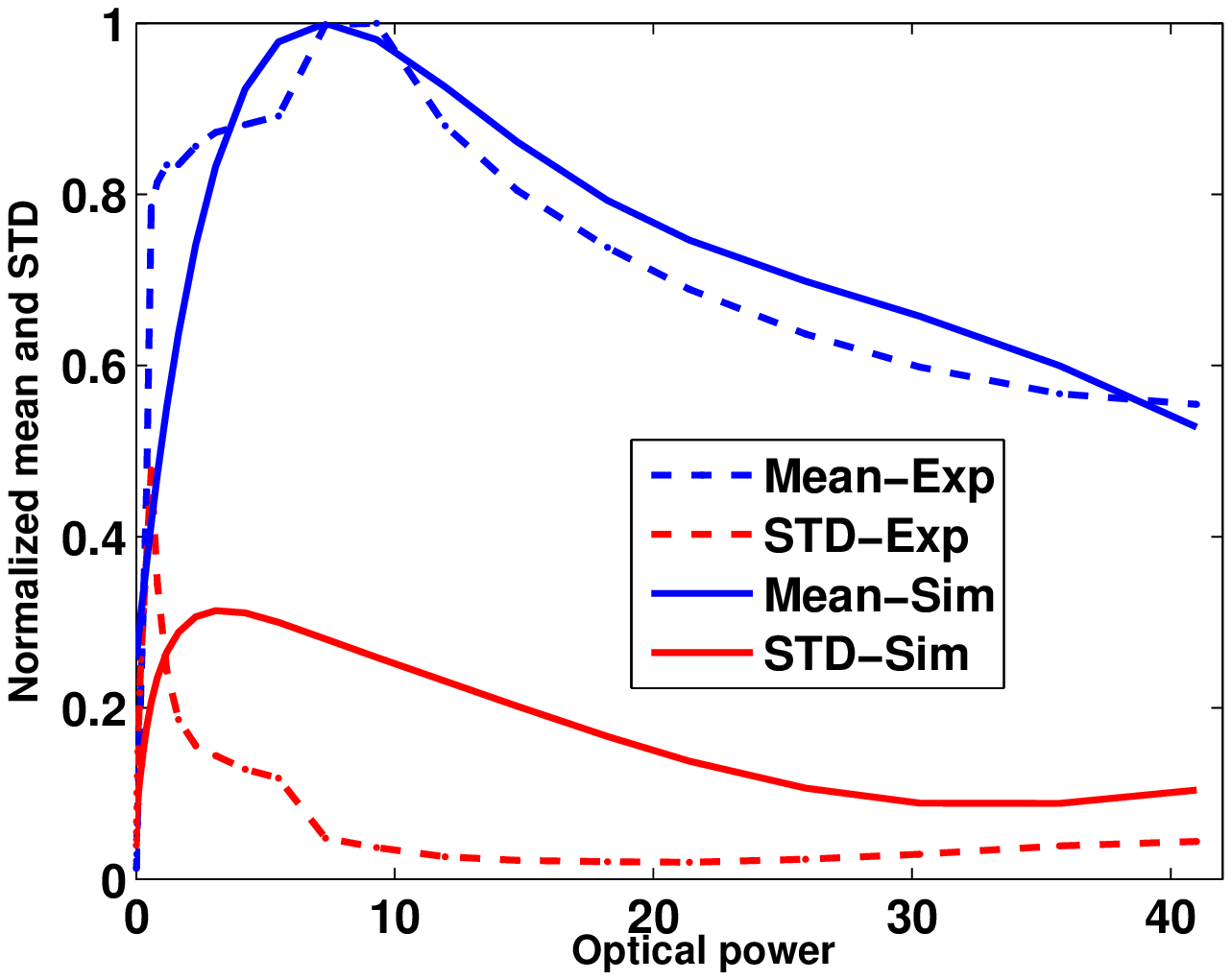}}
		\caption{The mean and STD of the received signals from experiments and simulation.}
		\label{fig.meanvar1}
	\end{minipage}
\end{figure}

\iffalse
\begin{figure}[htbp]
	\centering
	{\includegraphics[width =4.5in]{pdf-fitting.eps}}
	\caption{The PMT gain PDFs and the Gaussian approximations for different $\lambda$ values.}
	\label{fig.pdflam}
\end{figure}
\begin{figure}[htbp]
	\setlength{\abovecaptionskip}{0cm} %缩小caption和图像之间的距离
	\setlength{\belowcaptionskip}{0cm}
	\centering
	{\includegraphics[width =4.5in]{MPDover1.eps}}
	\caption{The error probabilities of MPD versus $\lambda_s$ for different over-sampling rates.}
	\label{fig.MPDover}
\end{figure}
\fi
\begin{figure}
		\centering
		{\includegraphics[angle=0, width=0.8\textwidth]{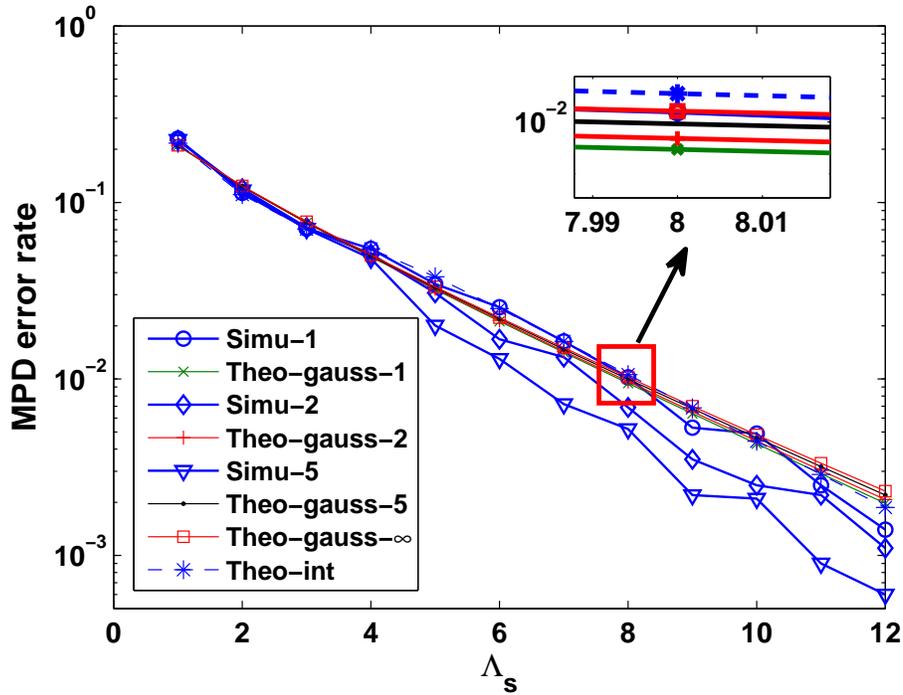}}
		\caption{The error probabilities of MPD versus $\lambda_s$ for different over-sampling rates.}
		\label{fig.MPDover}
\end{figure}
\begin{figure}
	\begin{minipage}[t]{0.45\textwidth}
		\centering
		{\includegraphics[angle=0, width=1.0\textwidth]{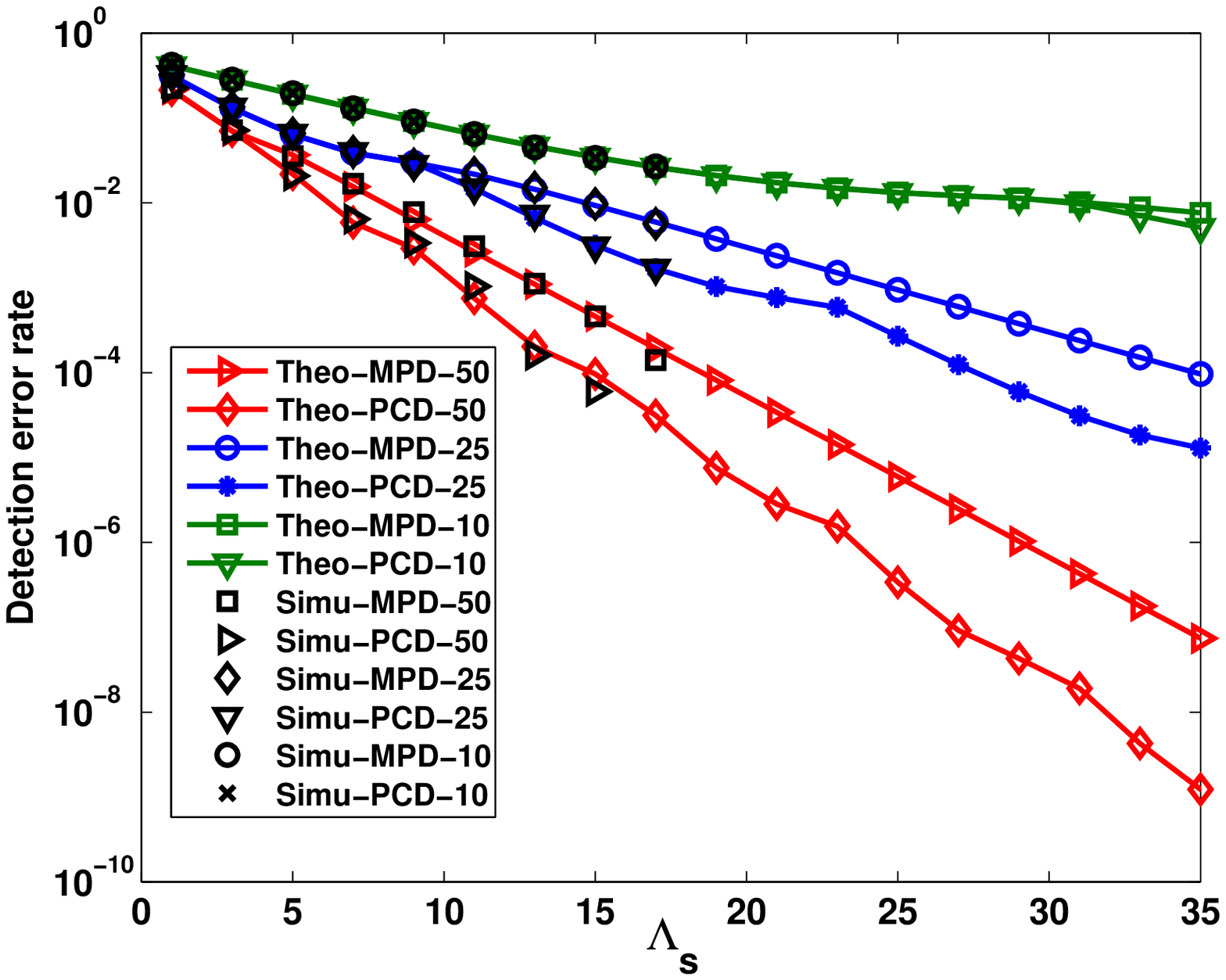}}
		\caption{The error rates of MPD and PCD versus $\lambda_s$ for different down-sampling rates in the linear regime.}
		\label{fig.BERlamdown}
	\end{minipage}
	\begin{minipage}[t]{0.45\textwidth}
		\centering
		{\includegraphics[angle=0, width=1.0\textwidth]{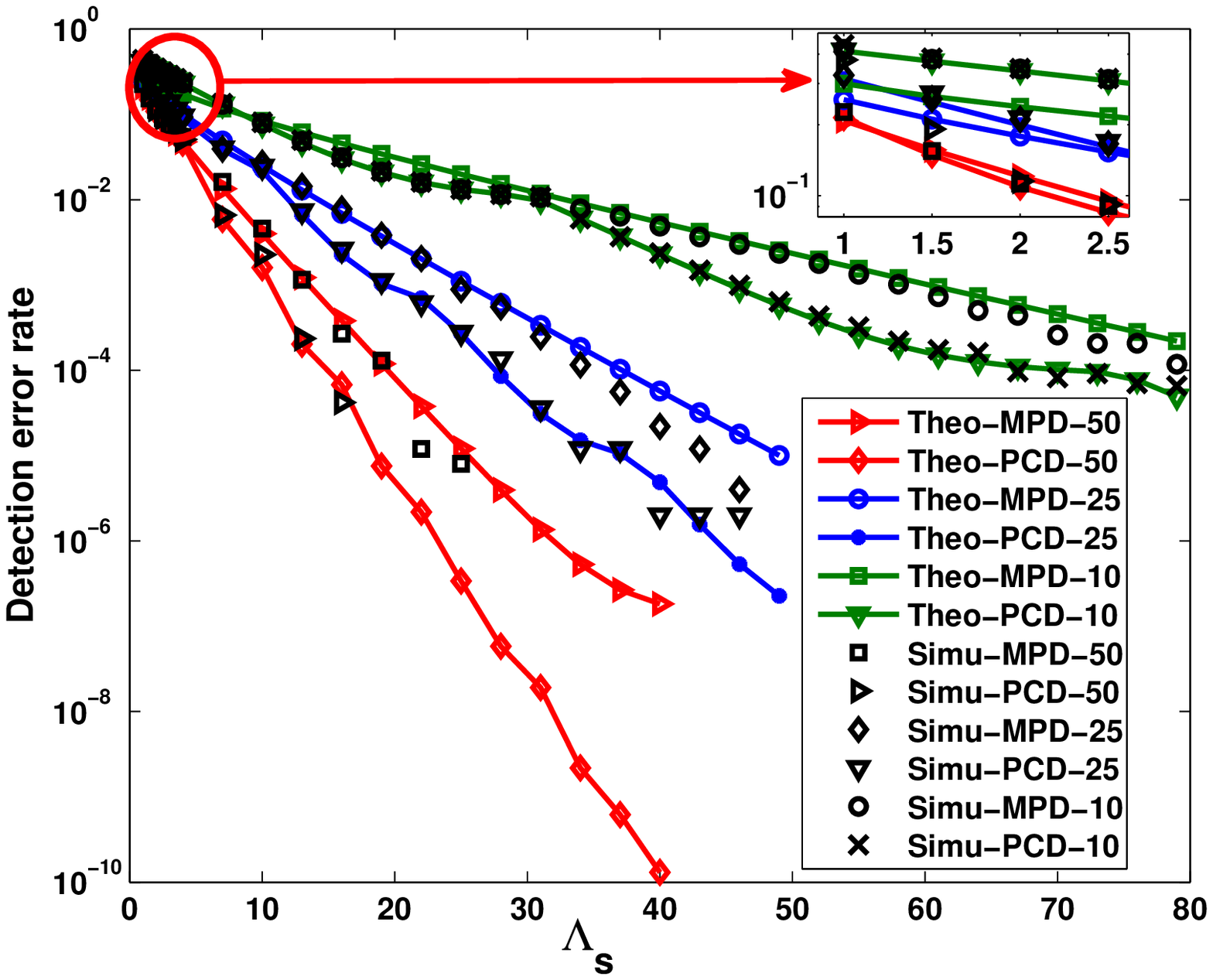}}
		\caption{The error rates of MPD and PCD versus $\lambda_s$ for different down-sampling rates in the non-linear regime.}
		\label{fig.BERlamdown2}
	\end{minipage}
\end{figure}
\iffalse
\begin{figure}[htbp]
	\centering
	{\includegraphics[width =4.5in]{AVPCD2.eps}}
	\caption{The error rates of MPD and PCD versus $\lambda_s$ for different down-sampling rates in linear case.}
	\label{fig.BERlamdown}
\end{figure}
\begin{figure}[htbp]
	\centering
	{\includegraphics[width =4.5in]{BERunder_nonli3.eps}}
	\caption{The error rates of MPD and PCD versus $\lambda_s$ for different down sample rates in non-linear case.}
	\label{fig.BERlamdown2}
\end{figure}
\fi

\section{conclusion}\label{sec.Conclusion}
We have studied the model on the PMT output signals based on multi-stage random amplification mechanism. We have also investigated the achievable rate for single and multiple sampling rates along with the optimal duty cycle and the suboptimal duty cycle based on small $\lambda_0$. Besides, we have proposed a threshold-based classifier to distinguish three working regimes of PMT. Moreover, we have investigated the performance of two signal detection approaches with different sampling rates, including the MPD and PCD.

\appendices
\section{Proof of achievable rate}
\subsection{Proof of Lemma \ref{lem.1}}\label{appd.lem1}
Defining $a_n(\tilde{z},\lambda)\dff\frac{e^{-a\tilde{z}}I_1(2a\sqrt{n\tilde{z}})}{\sqrt{n!}\sqrt{\tilde{z}}}$, $b_n(\tilde{z},\lambda)\dff a\frac{\sqrt{n}(\lambda e^{-a})^n}{\sqrt{n!}}e^{-\lambda}$ and sequence of functions ${f_n(\tilde{z},\lambda)=\sum_{i=0}^{n}a_i(\tilde{z},\lambda)b_i(\tilde{z},\lambda)}$, we have $\lim\limits_{n\to\infty}f_n(\tilde{z},\lambda)=f_{\tilde{Z}|\lambda}^{C}(\tilde{z})$. According to Equation (2.4) in \cite{ifantis1991bounds}, we have $e^y(\frac{y+3}{4})^4<\Gamma(2)\frac{2}{y}I_1(y)<e^y(\frac{3}{2y+3})^{\frac{3}{2}}$ for $y>0$. Thus  $a_n(\tilde{z},\lambda)<\frac{e^{2a\sqrt{n\tilde{z}}-a\tilde{z}}a\sqrt{n}}{\Gamma(2)\sqrt{n!}}\leq\frac{e^{an}a\sqrt{n}}{\Gamma(2)\sqrt{n!}}\rightarrow0$ for $n\to\infty$. Noting that there exists sufficiently large $N_1$ such that $\sqrt{(n-1)!}>e^{2an}$ for $n>N_1$, we have $\sum_{n=0}^{+\infty}a_n(\tilde{z},\lambda)<\sum_{n=0}^{+\infty}\frac{e^{an}a\sqrt{n}}{\Gamma(2)\sqrt{n!}}<\sum_{n=0}^{N_1}\frac{e^{an}a\sqrt{n}}{\Gamma(2)\sqrt{n!}}+\sum_{n=N_1+1}^{+\infty}\frac{e^{-an}}{\Gamma(2)}<\sum_{n=0}^{N_1}\frac{e^{an}a\sqrt{n}}{\Gamma(2)\sqrt{n!}}+\frac{e^{-aN_1}}{\Gamma(2)(1-e^{-aN_1})}$. Thus partial sum of $\sum_{n=0}^{+\infty}a_n(\tilde{z},\lambda)$ is uniformly bounded.  

Note that for fixed $\lambda$, $b_n$ attenuates to 0 as $n$ approaches infinity. According to Dirichlet's test, $\sum_{i=0}^{\infty}a_i(\tilde{z},\lambda)b_i(\tilde{z},\lambda)$ uniformly converges in $\mathbf{R}_+\times(0,\lambda_A]$. As each term of $a_i(\tilde{z},\lambda)b_i(\tilde{z},\lambda)$ is continuous, function $f_{\tilde{Z}|\lambda}^{C}(\tilde{z})$ is uniformly continuous
with respect to $(\tilde{z},\lambda)\in\mathbf{R}_+\times(0,\lambda_A]$.		
\subsection{Proof of Lemma \ref{lemma.epsilon}}\label{appd.lem2}
Since $f_{\tilde{Z}|\lambda_0}^{C}(\tilde{z})-\epsilon f_{\tilde{Z}|\lambda_1}^{C}(\tilde{z})=a\sqrt{\frac{1}{z}}e^{-az}\sum_{n=0}^{\infty}(e^{-\lambda_0}\lambda_0^n-\epsilon e^{-\lambda_1}\lambda_1^n)\frac{\sqrt{n}e^{-na}}{n!}I_1(2a\sqrt{nz})$, it suffices to have that $f_{\tilde{Z}|\lambda_0}^{C}(\tilde{z})<\epsilon f_{\tilde{Z}|\lambda_1}^{C}(\tilde{z})$ if $e^{-\lambda_0}\lambda_0^n<\epsilon e^{-\lambda_1}\lambda_1^n$ for any $n\in\mathbf{N}_{+}$. In the case of $\lambda_1\leq1$, for $\lambda_0<\min\limits_{n\in\mathbf{N}_{+}}\sqrt[n]{\frac{\epsilon}{e}}\lambda_1=\frac{\epsilon}{e}\lambda_1$ we have $e^{-\lambda_0}\lambda_0^n<\lambda_0^n<\frac{\epsilon}{e}\lambda_1^n\leq\epsilon e^{-\lambda_1}\lambda_1^n$ for any $n\in\mathbf{N}_{+}$. In the case of $\lambda_1>1$, for $\lambda_0<\min\limits_{n\in\mathbf{N}_{+}}\sqrt[n]{\epsilon e^{-\lambda_1}\lambda_1}=\epsilon e^{-\lambda_1}\lambda_1$, we have $e^{-\lambda_0}\lambda_0^n<\lambda_0^n<\epsilon e^{-\lambda_1}\lambda_1\leq\epsilon e^{-\lambda_1}\lambda_1^n$ for any $n\in\mathbf{N}_{+}$. In summary, defining $\delta=\epsilon e^{-\max\{1,\lambda_1\}}\lambda_1$, the condition for both $\lambda_1 \leq 1$ and $\lambda_1 > 1$ can be satisfied, and thus we complete the proof.	
\subsection{Proof of Lemma \ref{lemma.appmutual}}\label{appd.apprmutual}
Note that 
\be 
\log\big(\mu f_{\tilde{Z}|\lambda_1}^{C}(\tilde{z})+(1-\mu)f_{\tilde{Z}|\lambda_0}^{C}(\tilde{z})\big)&=&\log\big(\mu f_{\tilde{Z}|\lambda_1}^{C}(\tilde{z})\big)+\frac{1-\mu}{\mu}\frac{f_{\tilde{Z}|\lambda_0}^{C}(\tilde{z})}{f_{\tilde{Z}|\lambda_1}^{C}(\tilde{z})}+o(\lambda_0), \\
\int_{0}^{\infty}f_{\tilde{Z}|\lambda}^{C}(\tilde{z})\mathrm{d}\tilde{z}&=&1-e^{-\lambda(1-e^{-a})}.
\ee 
Based on Lemma \ref{lemma.epsilon}, for $\lambda_0<\delta=\epsilon e^{-\max\{1,\lambda_1\}}\lambda_1$, we have 
\be 
\int_{0}^{+\infty}\frac{[f_{\tilde{Z}|\lambda_0}^{C}(\tilde{z})]^2}{f_{\tilde{Z}|\lambda_1}^{C}(\tilde{z})}\mathrm{d}\tilde{z}<\epsilon\int_{0}^{+\infty}\{f_{\tilde{Z}|\lambda_0}^{C}(\tilde{z})\}\mathrm{d}\tilde{z}=\epsilon\big(1-e^{-\lambda_0(1-e^{-a})}\big),
\ee 
which implies that $\int_{0}^{+\infty}\frac{[f_{\tilde{Z}|\lambda_0}^{C}(\tilde{z})]^2}{f_{\tilde{Z}|\lambda_1}^{C}(\tilde{z})}\mathrm{d}\tilde{z}=o(\lambda_0)$. According to Cauchy-Schwarz inequality, we have the following
\be 
\int_{0}^{+\infty}\frac{[f_{\tilde{Z}|\lambda_0}^{C}(\tilde{z})]^2}{f_{\tilde{Z}|\lambda_1}^{C}(\tilde{z})}\mathrm{d}\tilde{z}\int_{0}^{+\infty}f_{\tilde{Z}|\lambda_1}^{C}(\tilde{z})\mathrm{d}\tilde{z}\geq\big(\int_{0}^{+\infty}f_{\tilde{Z}|\lambda_0}^{C}(\tilde{z})\mathrm{d}\tilde{z}\big)^2=\big(1-e^{-\lambda_0(1-e^{-a})}\big)^2,
\ee 
which implies that $\int_{0}^{+\infty}\frac{[f_{\tilde{Z}|\lambda_0}^{C}(\tilde{z})]^2}{f_{\tilde{Z}|\lambda_1}^{C}(\tilde{z})}\mathrm{d}\tilde{z}\geq C_0\lambda_0^2$ for $0<\lambda_0<\delta$ for certain constant $C_0>0$. 
Thus, the output signal entropy is given by
\be\label{eq.Hmix}
&&H\big(\mu f_{\tilde{Z}|\lambda_1}^{C}(\tilde{z})+(1-\mu)f_{\tilde{Z}|\lambda_0}^{C}(\tilde{z})\big)\\ \nonumber&=&H\big(\mu f_{\tilde{Z}|\lambda_1}^{C}\big)+H_2\big((1-\mu) f_{\tilde{Z}|\lambda_0}^{C},\mu f_{\tilde{Z}|\lambda_1}^{C}\big)-(1-\mu)\int_{0}^{+\infty}f_{\tilde{Z}|\lambda_0}^{C}(\tilde{z})\mathrm{d}\tilde{z}\\ \nonumber&&-\frac{(1-\mu)^2}{\mu}\int_{0}^{+\infty}\frac{[f_{\tilde{Z}|\lambda_0}^{C}(\tilde{z})]^2}{f_{\tilde{Z}|\lambda_1}^{C}(\tilde{z})}\mathrm{d}\tilde{z}+o(\lambda_0)\\ \nonumber
&=&\mu H\big(f_{\tilde{Z}|\lambda_1}^{C}\big)+(1-\mu)H_2\big(f_{\tilde{Z}|\lambda_0}^{C},f_{\tilde{Z}|\lambda_1}^{C}\big)-\mu\log\mu\big(1-(e^{-\lambda_1(1-e^{-a})})\big)\\ \nonumber
&&-(1-\mu)\log\mu\big(1-(e^{-\lambda_0(1-e^{-a})})\big)-(1-\mu)\big(1-(e^{-\lambda_0(1-e^{-a})})\big)+o(\lambda_0).
\ee 
Plugging Equation (\ref{eq.Hmix}) into Equation (\ref{eq.Icon}), we have
\be\label{eq.Icon2}
I^{C}(\xi;\tilde{Z})&=&(1-\mu)H_2\big(f_{\tilde{Z}|\lambda_0}^{C},f_{\tilde{Z}|\lambda_1}^{C}\big)-(1-\mu) H\big(f_{\tilde{Z}|\lambda_0}^{C}\big)-\mu\log\mu\big(1-(e^{-\lambda_1(1-e^{-a})})\big)\\ \nonumber
&&-(1-\mu)\log\mu\big(1-e^{-\lambda_0(1-e^{-a})}\big)-(1-\mu)\big(1-e^{-\lambda_0(1-e^{-a})}\big)+o(\lambda_0).
\ee 
Noting that $f_{\tilde{Z}|\lambda_0}(\tilde{z})=(a_0+o(\lambda_0))\delta(\tilde{z})+\big(\lambda_0+o(\lambda_0)\big)g(\tilde{z})$, we have
\be\label{eq.H_2}
H_2\big(f_{\tilde{Z}|\lambda_0}^{C},f_{\tilde{Z}|\lambda_1}^{C}\big)= \lambda_0 C_1+o(\lambda_0),
\ee 
where $C_1=-\int_{0}^{+\infty}g(\tilde{z})\log f_{\tilde{Z}|\lambda_1}^{C}(\tilde{z})\mathrm{d}\tilde{z}$ and
\be\label{eq.H_1}
H\big(f_{\tilde{Z}|\lambda_0}^{C}\big)&=&-(\lambda_0+o(\lambda_0))\int_{0}^{+\infty}g(\tilde{z})\log g(\tilde{z})\mathrm{d}\tilde{z}-(\lambda_0\log{\lambda_0}+o(\lambda_0))\int_{0}^{+\infty}g(\tilde{z})\mathrm{d}\tilde{z}\\ \nonumber
&=&\lambda_0C_2-\lambda_0\log\lambda_0(1-e^{-a})+o(\lambda_0),
\ee 
where $C_2=-\int_{0}^{+\infty}g(\tilde{z})\log g(\tilde{z})\mathrm{d}\tilde{z}$. Plugging Equations (\ref{eq.H_2}) and (\ref{eq.H_1}) into Equation (\ref{eq.Icon2}), we have
\be 
I^{C}(\xi;\tilde{Z})&=&-\mu\log\mu\big(1-(e^{-\lambda_1(1-e^{-a})})\big)-(1-\mu)\log\mu\lambda_0(1-e^{-a})\\ \nonumber
&&-(1-\mu)\lambda_0(1-e^{-a})+\lambda_0(1-\mu)(C_1-C_2)+o(\lambda_0).
\ee 
\subsection{Proof of Theorem \ref{theo.submu1}}\label{appd.theor1}
Defining $G(\mu)=-\mu\log\mu(1-a_1)-(\mu a_1+1-\mu)\log(\mu a_1+1-\mu)+\mu a_1\log a_1$ with $\mu\in[0,1]$, we have
\be\label{eq.submu1}
G^{'}(\mu)&=&(1-a_1)[\log(\mu a_1+1-\mu)-\log(\mu)]+a_1\log a_1;\\
G^{''}(\mu)&=&\frac{-(1-a_1)}{\mu(\mu a_1+1-\mu)}<0.
\ee	
Since $\lim\limits_{\mu\to0}G^{'}(\mu)=+\infty>0$ and $G^{'}(1)=\log a_1<0$, equation $G^{'}(\mu)=0$ has unique solution $\mu_a=\frac{1}{1-a_1+a_1^{-\frac{a_1}{1-a_1}}}$. 

From the above arguments, it is seen that $\max\limits_{0\leq\mu\leq\eta}G(\mu)$ is achieved for $\mu=\mu_a$, provided $\eta\geq\mu_a$. When $\mu_a\geq\eta$ (from the concavity of $G(\mu)$), the maximum is achieved for $\mu=\eta$. Thus, we have
\be 
\mu_1^{*}=\min\{\eta,\mu_a\}.
\ee 
Furthermore, since
\be
\lim\limits_{\lambda_A\to0}a_1^{-\frac{a_1}{1-a_1}}=\lim\limits_{a_1\to1}[(1+(a_1-1))^{\frac{1}{a_1-1}}]^{a_1}=e,
\lim\limits_{\lambda_A\to+\infty}a_1^{-\frac{a_1}{1-a_1}}=e^{\lim\limits_{a_1\to0}-\frac{a_1}{1-a_1}\ln{a_1}}=1,
\ee
we have $\lim\limits_{\lambda_A\to0}\mu_a=\frac{1}{e}$ and $\lim\limits_{\lambda_A\to+\infty}\mu_a=\frac{1}{2}$. 
\subsection{Proof of Theorem \ref{theor.optmu1}}\label{appd.theor2}
Define $G_2(\mu)\dff I(\xi,\tilde{Z})$ with $\mu\in[0,1]$. Based on Equation (\ref{eq.Idis}) and Equation (\ref{eq.Icon}), we have
\be
G_2^{'}(\mu)&=&\int_{0}^{+\infty}\big(f_{\tilde{Z}|\lambda_0}^{C}(\tilde{z})-  f_{\tilde{Z}|\lambda_1}^{C}(\tilde{z})\big)\{1+\log[\mu f_{\tilde{Z}|\lambda_1}^{C}(\tilde{z})+(1-\mu)f_{\tilde{Z}|\lambda_0}^{C}(\tilde{z})]\}\mathrm{d}\tilde{z}\\ \nonumber &&-H(f_{\tilde{Z}|\lambda_1}^{C})+H(f_{\tilde{Z}|\lambda_0}^{C})-(a_1-a_0)\{1+\log[\mu a_1+(1-\mu)a_0]\}+a_1\log a_1-a_0\log a_0.
\ee	
Noting that $\int_{0}^{+\infty}f_{\tilde{Z}|\lambda_0}^{C}(\tilde{z})- f_{\tilde{Z}|\lambda_1}^{C}(\tilde{z})\mathrm{d}\tilde{z}=a_1-a_0$, we have
\be 
G_2^{'}(\mu)&=&\int_{0}^{+\infty}\big(f_{\tilde{Z}|\lambda_0}^{C}(\tilde{z})-f_{\tilde{Z}|\lambda_1}^{C}(\tilde{z})\big)\log[\mu f_{\tilde{Z}|\lambda_1}^{C}(\tilde{z})+(1-\mu)f_{\tilde{Z}|\lambda_0}^{C}(\tilde{z})]\mathrm{d}\tilde{z}-H(f_{\tilde{Z}|\lambda_1}^{C})\\ \nonumber &&+H(f_{\tilde{Z}|\lambda_0}^{C})-(a_1-a_0)\log[\mu a_1+(1-\mu)a_0]+a_1\log a_1-a_0\log a_0,\\
G_2^{''}(\mu)&=&\int_{0}^{+\infty}-\frac{\big(f_{\tilde{Z}|\lambda_0}^{C}(\tilde{z})-f_{\tilde{Z}|\lambda_1}^{C}(\tilde{z})\big)^2}{\mu f_{\tilde{Z}|\lambda_1}^{C}(\tilde{z})+(1-\mu)f_{\tilde{Z}|\lambda_0}^{C}(\tilde{z})}\mathrm{d}\tilde{z}-\frac{(a_1-a_0)^2}{\mu a_1+(1-\mu)a_0}<0.
\ee 
Thus, there exists at most one solution to $G_2^{'}(\mu)=0$. The existence of such solution is justified as follows
\be
\lim\limits_{\mu\to0}G_2^{'}(\mu)&=&KL(f_{\tilde{Z}|\lambda_1}(\tilde{z}),f_{\tilde{Z}|\lambda_0}(\tilde{z}))>0, \\ 
\lim\limits_{\mu\to1}G_2^{'}(\mu)&=&-KL(f_{\tilde{Z}|\lambda_0}(\tilde{z}),f_{\tilde{Z}|\lambda_1}(\tilde{z}))<0.  
\ee 
From the above arguments we have that $\max\limits_{0\leq\mu\leq\eta}G_2(\mu)$ is achieved by $\mu_1^{*}=\mu_1$, for $\eta\geq\mu_1$. For $\mu_1\geq\eta$, this maximum value is achieved for $\mu_1^{*}=\eta$. Thus, we have
\be 
\mu_1^{*}=\min\{\eta,\mu_1\}.
\ee 
\subsection{Proof of Lemma \ref{lem.mutumulti}}\label{appd.mutumulti}
Define probability density function $f_{i|j}\dff f_{\tilde{z}_i|\xi=j}$. Since samples $\tilde{Z}^\Omega$ are mutually independent for $T_s\geq\tau$, we have
\be 
f_{\tilde{Z}^\Omega|j}(\tilde{z}^\Omega)=\prod_{i=1}^{L}[a_j\delta(\tilde{z}_i)+f_{i|j}(\tilde{z}_i)]=\sum\limits_{S\subseteq\Omega}a_j^{|S|}\delta(\tilde{z}^S)f_{S^C|j}^C(\tilde{z}^{S^C}),\quad j=0,1,
\ee
where $|S|$ denotes the cardinality of set $S$. Since
\be 
&&\int\log\frac{\mathrm{d}F_{\tilde{Z}^\Omega
		|j}(\tilde{z}^\Omega)}{\mathrm{d}F_{\tilde{Z}^\Omega}(\tilde{z}^\Omega)}\mathrm{d}F_{\tilde{Z}^\Omega|j}(\tilde{z}^\Omega)\\ \nonumber&=&\sum\limits_{S\subseteq\Omega}\int\log\frac{\mathrm{d}F_{\tilde{Z}^\Omega|j}(\tilde{z}^\Omega)}{\mathrm{d}F_{\tilde{Z}^\Omega}(\tilde{z}^\Omega)}a_j^{|S|}\delta(\tilde{z}^S)f_{S^C|j}^C(\tilde{z}^{S^C})\mathrm{d}\tilde{z}^\Omega\\ \nonumber
&=&\sum\limits_{S\subseteq\Omega}\int f_{S^C|j}^C(\tilde{z}^{S^C})a_j^{|S|}\log\frac{a_j^{|S|}f_{S^C|j}^{C}(z^{S^C})}{\mu a_1^{|S|}f_{S^C|1}^{C}(z^{S^C})+(1-\mu)a_0^{|S|}f_{S^C|0}^{C}(z^{S^C})}\mathrm{d}\tilde{z}^{S^C}\\ \nonumber
&=&\sum\limits_{S\subseteq\Omega}KL\big(a_j^{|S|}f_{S^C|j}^{C},\mu a_1^{|S|}f_{S^C|1}^{C}+(1-\mu)a_0^{|S|}f_{S^C|0}^{C}\big),
\ee 
we have the following on the mutual information
\be 
I(\xi;\tilde{Z}^\Omega)&=&\int\log\frac{\mathrm{d}F_{\xi,\tilde{Z}^\Omega}(\xi,\tilde{z}^\Omega)}{\mathrm{d}F_{\xi}(\xi)\mathrm{d}F_{\tilde{Z}^\Omega}(\tilde{z}^\Omega)}\mathrm{d}F_{\xi,\tilde{Z}^\Omega}(\xi,\tilde{z}^\Omega)\\ \nonumber
&=&\mu\int\log\frac{\mathrm{d}F_{\tilde{Z}^\Omega|1}(\tilde{z}^\Omega)}{\mathrm{d}F_{\tilde{Z}^\Omega}(\tilde{z}^\Omega)}\mathrm{d}F_{\tilde{Z}^\Omega|1}(\tilde{z}^\Omega)+(1-\mu)\int\log\frac{\mathrm{d}F_{\tilde{Z}^\Omega|0}(\tilde{z}^\Omega)}{\mathrm{d}F_{\tilde{Z}^\Omega}(\tilde{z}^\Omega)}\mathrm{d}F_{\tilde{Z}^\Omega|0}(\tilde{z}^\Omega)\\ \nonumber
&=&\sum\limits_{S\subseteq\Omega}\mu KL\big(a_1^{|S|}f_{S^C|1}^{C},\mu a_1^{|S|}f_{S^C|1}^{C}+(1-\mu)a_0^{|S|}f_{S^C|0}^{C}\big)\\ \nonumber &&+(1-\mu)KL\big(a_0^{|S|}f_{S^C|0}^{C},\mu a_1^{|S|}f_{S^C|1}^{C}+(1-\mu)a_0^{|S|}f_{S^C|0}^{C}\big) \\ \nonumber
&\dff&\sum\limits_{S\subseteq\Omega}I^S(\xi,\tilde{Z}),
\ee 
where $I^S(\xi,\tilde{Z})=H(\mu a_1^{|S|}f_{S^C|1}^{C}+(1-\mu)a_0^{|S|}f_{S^C|0}^{C})-\mu H(a_1^{|S|}f_{S^C|1}^{C})-(1-\mu)H(a_0^{|S|}f_{S^C|0}^{C})$ (define $f_{\o|j}^{C}=1$ and $H(a_j^{|\Omega|}f_{\o|j}^{C})=-a_j^{|\Omega|}\log(a_j^{|\Omega|})$).
\subsection{Proof of Theorem \ref{theor.3}}\label{appd.theor3}
Firstly, we show that $I(\xi;\tilde{Z}^\Omega)$ is concave with respect to $\mu$. Since
\be 
\frac{\partial I^S}{\partial\mu}&=&\int_{\mathbf{R}_{+}^{|\Omega|-|S|}}\big(a_0^{|S|}f_{S^C|0}^C(\tilde{z}^{S^C})-a_1^{|S|}f_{S^C|1}^C(\tilde{z}^{S^C})\big)\log\big(\mu a_1^{|S|}f_{S^C|1}^{C}(\tilde{z}^{S^C})\\ \nonumber
&&+(1-\mu)a_0^{|S|}f_{S^C|0}^{C}(\tilde{z}^{S^C})\big)\mathrm{d}\tilde{z}^{S^C}-H(a_1^{|S|}f_{S^C|1}^C)+H(a_0^{|S|}f_{S^C|0}^C), \\
\frac{\partial^2 I^S}{\partial\mu^2}&=&\int_{\mathbf{R}_{+}^{|\Omega|-|S|}}-\frac{\big(a_0^{|S|}f_{S^C|0}^C(\tilde{z}^{S^C})-a_1^{|S|}f_{S^C|1}^C(\tilde{z}^{S^C})\big)^2}{\mu a_1^{|S|}f_{S^C|1}^{C}(\tilde{z}^{S^C})
	+(1-\mu)a_0^{|S|}f_{S^C|0}^{C}(\tilde{z}^{S^C})}\mathrm{d}\tilde{z}^{S^C}<0,
\ee
we have $\frac{\partial^2 I}{\partial\mu^2}=\sum\limits_{S\subseteq\Omega}\frac{\partial^2 I^S}{\partial\mu^2}<0$. Then, there exists at most one solution to $\frac{\partial I}{\partial\mu}=0$. Noting 
\be 
&&-\sum\limits_{S\subseteq\Omega}\int_{\mathbf{R}_{+}^{|\Omega|-|S|}}a_0^{|S|}f_{S^C|0}^C(\tilde{z}^{S^C})-a_1^{|S|}f_{S^C|1}^C(\tilde{z}^{S^C})\mathrm{d}\tilde{z}^{S^C}\\ \nonumber&=&-\sum\limits_{S\subseteq\Omega}a_0^{|S|}(1-a_0)^{|\Omega|-|S|}-a_1^{|S|}(1-a_1)^{|\Omega|-|S|}\\ \nonumber&=&-[(a_0+1-a_0)]^{|\Omega|}-(a_1+1-a_1)]^{|\Omega|}]=0,
\ee
we have
\be 
\frac{\partial I}{\partial\mu}&=&\sum\limits_{S\subseteq\Omega}\int_{\mathbf{R}_{+}^{|\Omega|-|S|}}\big(a_0^{|S|}f_{S^C|0}^C(\tilde{z}^{S^C})-a_1^{|S|}f_{S^C|1}^C(\tilde{z}^{S^C})\big)\log\big(\mu a_1^{|S|}f_{S^C|1}^{C}(\tilde{z}^{S^C})\\ \nonumber
&&+(1-\mu)a_0^{|S|}f_{S^C|0}^{C}(\tilde{z}^{S^C})\big)\mathrm{d}\tilde{z}^{S^C}-H(a_1^{|S|}f_{S^C|1}^C)+H(a_0^{|S|}f_{S^C|0}^C).
\ee
The existence of solution is as follows
\be
\lim\limits_{\mu\to0}\frac{\partial I}{\partial\mu}&=&\sum\limits_{S\subseteq\Omega}\int_{\mathbf{R}_{+}^{|\Omega|-|S|}}a_1^{|S|}f_{S^C|1}^{C}(\tilde{z}^{S^C})\log\frac{a_1^{|S|}f_{S^C|1}^{C}(\tilde{z}^{S^C})}{a_0^{|S|}f_{S^C|0}^{C}(\tilde{z}^{S^C})}\mathrm{d}\tilde{z}^{S^C}\\ \nonumber
&\overset{(a)}{>}&\sum\limits_{S\subseteq\Omega}\int_{\mathbf{R}_{+}^{|\Omega|-|S|}}a_1^{|S|}f_{S^C|1}^{C}(\tilde{z}^{S^C})\mathrm{d}\tilde{z}^{S^C}\log\frac{\sum\limits_{S\subseteq\Omega}\int_{\mathbf{R}_{+}^{|\Omega|-|S|}}a_1^{|S|}f_{S^C|1}^{C}(\tilde{z}^{S^C})\mathrm{d}\tilde{z}^{S^C}}{\sum\limits_{S\subseteq\Omega}\int_{\mathbf{R}_{+}^{|\Omega|-|S|}}a_0^{|S|}f_{S^C|0}^{C}(\tilde{z}^{S^C})\mathrm{d}\tilde{z}^{S^C}}=0; \\  
\lim\limits_{\mu\to1}\frac{\partial I}{\partial\mu}&=&\sum\limits_{S\subseteq\Omega}\int_{\mathbf{R}_{+}^{|\Omega|-|S|}}-a_0^{|S|}f_{S^C|0}^{C}(\tilde{z}^{S^C})\log\frac{a_0^{|S|}f_{S^C|0}^{C}(\tilde{z}^{S^C})}{a_1^{|S|}f_{S^C|1}^{C}(\tilde{z}^{S^C})}\mathrm{d}\tilde{z}^{S^C}\overset{(b)}{<}0;
\ee 
where $(a)$ and $(b)$ follow from log-sum inequality and $\frac{f_{S^C|1}^{C}(\tilde{z}^{S^C})}{f_{S^C|0}^{C}(\tilde{z}^{S^C})}\neq const$. From the above statements we conclude that $\max\limits_{0\leq\mu\leq\eta}I(\xi;\tilde{Z}^{\Omega})$ is achieved for $\mu_m^{*}=\mu_m$ provided that $\eta\geq\mu_m$. When $\mu_m\geq\eta$, from the concavity of $I(\xi;\tilde{Z}^{\Omega})$ we have that the maximum is achieved for $\mu_m^{*}=\eta$. Thus, we have
\be 
\mu_m^{*}=\min\{\eta,\mu_m\}.
\ee 
\subsection{Proof of Theorem \ref{theor.4}}\label{appd.theor4}
Setting $\lambda_0=0$, we have $f_{S|0}(\tilde{z}^S)=\delta(\tilde{z}^S)$, and 
\be 
I^S(\xi,\tilde{Z})=H(\mu a_1^{|S|}f_{S^C|1}^{C}+(1-\mu)\mathbbm{1}\{S=\Omega\})-\mu H(a_1^{|S|}f_{S^C|1}^{C}).
\ee 
Then, we have
\be
\frac{\partial I^S}{\partial\mu}&=&\int_{\mathbf{R}_{+}^{L-|S|}}\big(\mathbbm
{1}\{S=\Omega\}-a_1^{|S|}f_{S^C|1}^C(\tilde{z}^{S^C})\big)\log\big(\mu a_1^{|S|}f_{S^C|1}^{C}(\tilde{z}^{S^C})\\ \nonumber
&&+(1-\mu)\mathbbm{1}\{S=\Omega\}\big)\mathrm{d}\tilde{z}^{S^C}-H(a_1^{|S|}f_{S^C|1}^C);\\ \label{eq.submu2} 
\frac{\partial I}{\partial\mu}&=&\sum\limits_{S\subsetneqq\Omega}-a_1^{|S|}(1-a-1)^{L-|S|}+(1-a_1^{L})\log(\mu a_1^{L}+1-\mu)+a_1^{L}\log a_1^{L} \\ \nonumber
&=&(1-a_1^{L})\log( a_1^{L}-1+\frac{1}{\mu})+a_1^{L}\log a_1^{L}.
\ee	
According to Equations (\ref{eq.submu1}) and (\ref{eq.submu2}), we can complete proof using similar procedures replacing $a_1$ with $a_1^{L}$ as that of Theorem \ref{theo.submu1}.
\section{Proof of working regime classification }
\subsection{Proof of Lemma \ref{lemma.exist}}\label{appd.lemexist}
Note that the contribution to $\tilde{Z}$ of each photon is independent and identically distributed. Based on Equation (\ref{eq.MGFG}), we have $\mathbb{E}[\tilde{G}]=1$ and $\mathbb{D}[\tilde{G}]=2a^{-1}$, where $\tilde{G}$ denotes the normalized PMT gain by $A$. Thus, we have
\be \mathbb{E}[\tilde{Z}|\Lambda]&=&\mathbb{E}[\mathbb{E}[\tilde{Z}|N,\Lambda]]=\mathbb{E}[N]=\lambda, \\ \mathbb{D}[\tilde{Z}|\Lambda]&=&\mathbb{D}[\mathbb{E}[\tilde{Z}|N,\Lambda]]+\mathbb{E}[\mathbb{D}[\tilde{Z}|N,\Lambda]]=\mathbb{D}[N]+\mathbb{E}[2a^{-1}N]=\lambda(1+2a^{-1}).
\ee
According to Chebshev inequality, we have 
\be 
\mathbb{P}(\tilde{Z}\leq l_{max}|\Lambda)\leq\frac{\Lambda(1+2a^{-1})}{(\lambda-l_{max})^2}\overset{\lambda\to\infty}{\longrightarrow}0.
\ee
Then for any $\epsilon>0$, there is a $\Lambda_d$ such that $\mathbb{P}(\tilde{Z}\leq l_{max}|\Lambda)<\epsilon$ for all $\Lambda\geq\Lambda_d$. The uniqueness is obvious.
\subsection{Proof of Lemma \ref{lem.gauap}}\label{appd.lemgauapp}
This proof is based on MGF. According to the translation property, we have
\be 
M_{\tilde{Z}_{nor}}(\mu|\Lambda)=e^{\sqrt{\frac{\lambda}{1+2a^{-1}}}\mu+\lambda\big(\exp(-\frac{\mu}{\sqrt{\lambda(1+2a^{-1})}+a^{-1}\mu})-1\big)}.
\ee
Since $-\frac{1}{x+a}=-\frac{1}{x}+\frac{a}{x(x+a)}$, according to Taylor's theorem, we have
\be 
&&\lambda\big(\exp(-\frac{\mu}{\sqrt{\lambda(1+2a^{-1})}+a^{-1}\mu})-1\big)\\ \nonumber&=&\lambda\Big(-\frac{\mu}{\sqrt{\lambda(1+2a^{-1})}+a^{-1}\mu}+(\frac{\mu}{\sqrt{\lambda(1+2a^{-1})}+a^{-1}\mu})^2+o(\frac{1}{\lambda})\Big)\\ \nonumber
&=&-\mu\sqrt{\frac{\lambda}{1+2a^{-1}}}+\frac{1}{2}\mu^2+o(1),
\ee 
where $o(1)$ means that $\lim\limits_{\Lambda\to\infty}o(1)=0$. Thus we have
\be 
\lim\limits_{\Lambda\to\infty}M_{\tilde{Z}_{nor}}(\mu|\Lambda)=e^{\frac{1}{2}\mu^2},
\ee
which is the MGF of $\mathcal{N}(0,1)$. According to L$\acute{e}$vy's continuity theorem \cite{williams1991probability}, we have that $\tilde{Z}_{nor}$ converges in distribution to $\mathcal{N}(0,1)$.
\subsection{Proof of Theorem \ref{theor.5}}\label{appd.theor5}
Since $\epsilon$ is small, typically $10^{-3}\sim10^{-2}$, the solution $\Lambda^{(2)}_{th}$ is large such that Gaussian approximation works well. According to Lemma \ref{lem.gauap}, we have the following on the Gaussian approximation,
\be 
\mathbb{P}(\tilde{Z}\leq l_{max}|\Lambda)\approx\Phi(\frac{l_{max}-\lambda}{\sqrt{\lambda(1+2a^{-1})}})\leq\epsilon,
\ee 
which leads to $l_{max}-\lambda\leq\Phi^{-1}(\epsilon)\sqrt{\lambda(1+2a^{-1})}$. Via straightforward mathematical derivations, we have 
\be 
\lambda^{(2)}_{th}=\frac{1}{4}\big(-\Phi^{-1}(\epsilon)\sqrt{1+2a^{-1}}+\sqrt{(1+2a^{-1})[\Phi^{-1}(\epsilon)]^2+4l_{max}}\big)^2.
\ee
\section{Proof of Mean Power Detection for Infinite Sampling Rate }
\subsection{Proof of Lemma \ref{lemma.meanpower}}\label{appd.meanpower}
According to Equation (\ref{eq.MGFG}), the conditional MGF of $a_j(t_k^j)$ given $t_k^j=\rho$, denoted as $M_{a_j|\rho}(\omega)$,
is given by 
\be 
M_{a_0|\rho}(\omega)=\mathbb{E}[\exp(-\omega a_0(t_k^0))|t_k^0=\rho]=\exp[-\frac{A\omega \alpha(\rho)}{1+B\omega \alpha(\rho)}],0\leq \rho <1;\\ \nonumber
M_{a_j|\rho}(\omega)=\mathbb{E}[\exp(-\omega a_j(t_k^j))|t_k^j=\rho]=\exp[-\frac{A\omega \alpha(\rho)}{1+B\omega \alpha(\rho)}],-j\leq \rho <-j+1.
\ee 
Note that $a_j(t_k^j)$ and $a_n(t_m^n)$ are independent for $k\neq m$ or $j\neq n$, and thus
\be
\prod_{k=1}^{m}\mathbb{E}[\exp(\omega a_j(t_k^j))|N=m]=\{\int_{-j}^{-j+1}M_{a_j|\rho}(\omega)d\rho\}^m.
\ee
Then, we have the following MGF of $r_j$
\be\label{eq.MGF_R}
M_{r_j|\Lambda}(\omega )&=&\mathbb{P}\{N=0\}+\sum_{m=1}^{\infty}\mathbb{P}\{N=m\}\prod_{k=1}^{m}\mathbb{E}[\exp(\omega a_j(t_k^j))|N=m]\\ \nonumber&=&\exp\Big(\Lambda[\int_{-j}^{-j+1}M_{a_j|\rho}(\omega)d\rho-1]\Big),\\ \nonumber
M_{r_j}(\omega)&=&\frac{1}{2}\Big(M_{r_j|\Lambda_s+\Lambda_0}(\omega)+M_{r_j|\Lambda_0}(\omega)\Big).
\ee 
Since $r_j$ and $r_k$ are independent for $j\neq k$, we have the following conditional MGF $M_{y_s|\xi}(\omega)$ of $y_s$ as equation (\ref{eq.MGF_Y}).		
\subsection{Proof of Corollary \ref{corol.meanpower2}}\label{appd.meanpower2}
Due to that $h(t)=u(t)-u(t-\tau)$, then we have
\be 
\alpha(\rho)=\left\{\begin{array}{ll}
	\tau,&0\leq\rho<1-\tau;\\
	1-\rho,&1-\tau\leq\rho<1;\\
	\tau+\rho,&-\tau\leq\rho<0;\\
	0,&\mathrm{otherwise}.
\end{array}\right.
\ee 
Based on Equation (\ref{eq.MGF_R}) and uniform distribution of photon arrival for homogeneous Poisson process, we have
\be 
M_{r_0|\Lambda}(\omega)&=&\exp\big(\Lambda[(1-\tau)e^{-\frac{\omega\tau}{1+a^{-1}\omega\tau}}+\int_{1-\tau}^{1}e^{-\frac{\omega(1-\rho)}{1+a^{-1}\omega(1-\rho)}}\mathrm{d}\rho-1]\big), \\
M_{r_1}(\omega|\Lambda)&=&\exp\big(\Lambda[(1-\tau)+\int_{0}^{\tau}e^{-\frac{\omega\rho}{1+a^{-1}\omega\rho}}\mathrm{d}\rho-1]\big), \\
M_{r_j|\Lambda}(\omega)&=&1,  \qquad j=2,3,\cdots.
\ee

Assuming that $\tau<<1$ for reliable communication, we resort to Taylor theorem:
\be 
\int_{1-\tau}^{1}e^{-\frac{\omega(1-\rho)}{1+a^{-1}\omega(1-\rho)}}\mathrm{d}\rho=\int_{0}^{\tau}e^{-\frac{\omega\rho}{1+a^{-1}\omega\rho}}\mathrm{d}\rho
=\tau-\frac{\omega}{2}\tau^2+o(\tau^2),
\ee 
which leads to the following,
\be 
M_{r_0|\Lambda}(\omega)&=&\exp[-\Lambda[(1-\tau)(1-e^{-\frac{\omega\tau}{1+a^{-1}\omega\tau}})-\frac{\omega}{2}\tau^2]+o(\tau^2),\\
M_{r_1|\Lambda}(\omega)&=&\exp(-\frac{\omega}{2}\tau^2\Lambda)+o(\tau^2).
\ee 
Based on Lemma \ref{lemma.meanpower}, the proof of Corollary \ref{corol.meanpower2} completes.

\section{Non-linear function determination}\label{appd.deter}
Due to unmeasured inner current of anode in PMT, we adopt the mean and variance of the output samples with respect to optical power $\Lambda$, denoted as $g_m(\Lambda)$ and $g_v(\Lambda)$, respectively, from experiment to estimate the non-linear function. We aim to find non-linear function $C(x)$ such that $\int_{0}^{+\infty}C(x)f_{Z|\Lambda}(x)\mathrm{d}x$ and $\sqrt{\int_{0}^{+\infty}C^2(x)f_{Z|\Lambda}(x)\mathrm{d}x}$ can be well approximated by $g_m(\Lambda)$ and $\sqrt{g_m^2(\Lambda)+g_v(\Lambda)}$, respectively.

Considering finite experimental data, we adopt discrete optimization method. Denote $\{\Lambda_i,i=1,\cdots, M_1\}$ and $\{x_j,j=1,\cdots, M_2\}$ $(x_0=0)$ as the measured optical power set in the experiment and discrete variable set, respectively. Denote probability matrix as $\mathbf{P}\dff[p_{ij}]_{M_1\times M_2}$, where $p_{ij}=\int_{x_{j-1}}^{x_j}f_{Z|\Lambda_i}(x)\mathrm{d}x$ for $j>1$ and $p_{i1}=e^{\Lambda_i\tau(e^{-a}-1)}$ for $i=1,\cdots M_1$, $c\dff[c_1,\cdots,c_{M_2}]^T$, where $c_j=C(x_j)$ for $j=1,\cdots,M_2$, $g_1\dff[g_m(\Lambda_1),\cdots,g_m(\Lambda_{M1})]^T$ and $g_2\dff[\sqrt{g_m^2(\Lambda_1)+g_v(\Lambda_1)},\cdots,\sqrt{g_m^2(\Lambda_{M1})+g_v(\Lambda_{M1})}]^T$. Mean square loss function is employed to measure the distance between the measured value and ideal value. For simplicity, letting $\cdot^2$ and $\sqrt{\cdot}$ denote the square and square root in the element-wise manner, respectively. The objective function and its gradient are given by
	\be 
	F(c)&=&||\mathbf{P}c-g_1||^2+||\sqrt{\mathbf{P}c^2}-g_2||^2,\\
	\frac{\partial F}{\partial c}&=&2\mathbf{P}^T(\mathbf{P}c-g_1)+2d,
	\ee	
	where $d\dff[d_1.\cdots,d_{M_2}]^T$ and $d_i=\sum_{i=1}^{M_1}\big(\sqrt{\sum_{j=1}^{M_2}p_{ij}c_j^2}-\sqrt{g_m^2(\Lambda_i)+g_v(\Lambda_i)}\big)\frac{p_{ij}c_j}{\sqrt{\sum_{j=1}^{M_2}p_{ij}c_j^2}}$. Noting that $M_2$ is typically large, we adopt Broyden-Fletcher-Goldfarb-Shanno (BFGS) algorithm for large scale unconstrained optimization problem \cite{nazareth1979relationship} for the fitting.

\small{\baselineskip = 10pt
\bibliographystyle{IEEEtran}
\bibliography{./pmt}

\end{document}